\documentclass[preprint,prc,aps,eqsecnum,epsf,amssymb,floatfix]{revtex4}
\usepackage{graphicx}

\newcommand{\qq}{{|\mbox{\boldmath $q$}|}}

\begin{document}
\preprint {WIS '10, Jan 1, DPPA}
\date{\today}
\title{Extraction of neutron observables from inclusive lepton scattering on nuclei}
\author{A.S. Rinat and M.F. Taragin}
\address{Weizmann Institute of Science, Department of Particle Physics,
Rehovot 76100, Israel}

\begin{abstract}

We analyze new JLAB data  for inclusive electron scattering on
various targets. Computed and measured total inclusive cross
sections in the range $0.3\lesssim x\lesssim 0.95$ show on a
logarithmic scale reasonable agreement for all targets. However,
closer inspection of the Quasi-Elastic components bares serious
discrepancies. EMC ratios with conceivably smaller systematic errors
fare the same. As a consequence the new data do not enable the
extraction of the magnetic form factor (FF) $G_M^n$ and the
Structure Function (SFs) $F_2^n$ of the neutron, although the
application of exactly the same analysis to older data had been
successful. We incorporate in the above analysis older CLAS
collaboration data on $F_2^D$. Removing some scattered points from
those, it appears possible to obtain the requested neutron
information. We compare our results with others from alternative
sources. Special attention is paid to the $A=3$ iso-doublet cross
sections and EMC ratios. Present data exist only for $^3$He, but the
available input in combination with charge symmetry enables
computations for $^3$H. Their average is the computed iso-scalar
part and is compared with the empirical modification of $^3$He EMC
ratios towards a fictitious $A=3$ iso-singlet.

\end{abstract}

\maketitle

\section{Introduction.}

Nearly a decade has passed since the publication of JLab experiment
E89-008, describing inclusive scattering of electrons on various
targets \cite{nicu1,arrd}. Those extended older SLAC data on the D
and He isotopes \cite{rock} and later ones

for beam energies $E\leq 3.6$ GeV on D, $^4$He and several medium
and heavy targets \cite{ne3}. A similar experiment in 1999 used
$E$=4.05 GeV electrons \cite{arr89}.

In the JLab experiments E03-102, E02-90, $5.76$ GeV unpolarized
electrons were scattered over angles $\theta=18^{\circ}, 22^{\circ},
26^{\circ}, 32^{\circ}, 40^{\circ}$ and $50^{\circ}$. Total
inclusive cross sections covering wide kinematics, have been
measured for targets D, $^{3,4}$He, $^9$Be, C, Al, Cu. Out of this
extensive data bank, only the EMC ratios $\mu^{^{3,4}He}, \mu^{Be}$
and $\mu^{C}$ for one scattering angle $\theta= 40^{\circ}$ and
limited kinematics have been published until this day \cite{seely}.
In addition cross section data over the entire measured kinematic
range and for all targets have been made available \cite{ag}. We
also mention data taken with a $E=5.0$ GeV beam, the analysis of
which has not yet been completed \cite{ps}.

In order to define notation, we start with the total cross section
per nucleon for inclusive scattering  of unpolarized electrons,
reduced by the Mott cross section $\sigma_M$. For given beam energy
$E$, scattering angle $\theta$ and energy loss $\nu$ one has
\begin{eqnarray}
K^A(x,Q^2,\theta)
&=&\frac{d^2\sigma^A(E;\theta,\nu)}{d\Omega\,d\nu}\Big/
     \sigma_M(E;\theta,\nu)\\
\nonumber
 &=&\frac {2xM}{Q^2} F_2^A(x,Q^2)+\frac{2}{M}F_1^A(x,Q^2){\rm
 tan}^2(\theta/2),
 \label{a1}
\end{eqnarray}
$F_{1,2}^A(x,Q^2)$ above are nuclear SFs, which depend on the
squared 4-momentum transfer $q^2=-Q^2=-(\qq^2-\nu^2)$ and the
Bjorken variable $0 \le x=Q^2/2M\nu \le M_A/M\approx A$ with $M$,
the nucleon mass.

Several approaches have been proposed for an analysis of inclusive
cross sections in the PWIA, or the same with some FSI distortions
\cite{benh,oset}. We report below on an analysis, based on a
previously tested non-perturbative GRS approach \cite{asr1,gr}. Our
application covers the entire corpus of new data (ND) also beyond
the restricted targets and kinematics of the published material,
reported in Refs. \cite{seely,ag}. To those we add an analysis of D
CLAS data \cite{osipd}. Except for the treatment of the $A=3$
targets, the method of analysis for the new data is identical to the
one, previously applied to the older data (OD). We therefore shall
not detail steps, but mention References instead.

We adhere in the following to a generalized convolution, linking
$F_k^A$ and $F_k^{p,n}$ (see for instance Ref. \cite{asr1})
\begin{eqnarray}
 F^A_k(x,Q^2)&=&\int_x^A\frac {dz}{z^{2-k}} \bigg [f_p^A(z,Q^2)
   ZF_k^p \bigg (\frac {x}{z},Q^2\bigg )
  +f_n^A(z,Q^2)NF_k^n\bigg (\frac {x}{z},Q^2\bigg ) \bigg ]\bigg/A.
  \\
  \label{a2a}
  &\approx&\int_x^A\frac {dz}{z^{2-k}} f^A(z,Q^2)
  \bigg [ZF_k^p \bigg (\frac {x}{z},Q^2\bigg )
  +NF_k^n\bigg (\frac {x}{z},Q^2\bigg ) \bigg ]/A.
\label{a2b}
\end{eqnarray}
Above $f^A$ are SFs for a fictitious nucleus composed of
point-particles which cannot be excited, irrespective of the value
of $Q^2$ \cite{aku}. Alternatively one interprets $f$ as a kind of
generalized distribution function of the centers of interacting
nucleons in a target.

The functions $f^A$ for finite $Q^2$ can only be calculated exactly
for the lightest nuclei and have otherwise to be modeled
\cite{asr1}. In the PWIA the above norm can be shown to be 1. The
same is expected for any bona-fide distribution function. We shall
return to this point in detail.

In virtually all previous applications one did not distinguish
between distribution functions $f_{p,n}^A$, which are different for
$p$ and $n$. However, in a treatment of the lightest odd nuclei,
their difference may matter and a proper treatment ought to use Eq.
(1.2).

Eqs. (1.2), (1.3) feature nucleon SFs $F_k^{p,n}$, which in general
are off their mass shell. However, in the region of our main
$Q^2\gtrsim (2.5-3.0)$ GeV$^2$, those effects may be neglected, and
the same holds for the mixing of nucleon SFs in the proper
expression for $F_2^A$ \cite{atw,sss}. Since the data do not reach
the deepest inelastic range $x\lesssim 0.2$, screening effects may
also be disregarded \cite{kul}. For $Q^2 \lesssim 3.0$ GeV$^2$, for
which (pseudo-)resonance structure is not yet extinguished, we shall
use $F_2^p$ from Ref. \cite{chr1}, while for larger $Q^2$ we rely on
a parametrization of the resonance-averaged $F_2^p$ \cite{arneo}.
References to $F_2^n$ can be found in \cite{asr1}.

It is convenient to decompose nucleon SFs $F_k^N$ in Eq. (1.2) into
parts, describing the absorption of a virtual photon, either
exciting the absorbing $N$ into hadrons (partons) or not ($\gamma^*+
N\to N$). The amplitudes for the latter vanish except for $x=1$, in
which case those may be  expressed as standard combinations of
electro-magnetic FFs. A similar division applies to nuclear SFs.
Denoting by $[{\tilde G}^N]^2=[Z(G^p)^2+N(G^n)^2]/A\,\,$, the $Z,N$
weighted average of the squared nucleon FFs, one finds from Eq.
(\ref{a2a}) their nuclear analogs ($\eta=Q^2/(4M^2)$)
\begin{mathletters}
\begin{eqnarray}
  F_1^{A,NE}(x,Q^2)&=& \frac {f^{PN,A}(x,Q^2)}{2}[{\tilde G}_M^N(Q^2)]^2] \,
  \label{a3a}\\
  F_2^{A,NE}(x,Q^2)&=&xf^{PN,A}(x,Q^2) \frac{[{\tilde G}_E^N(Q^2)]^2
 +\eta [{\tilde G}_M^N(Q^2)]^2}{1+\eta}
  \label{a3b}
\end{eqnarray}
\end{mathletters}
Nuclear inelastic processes (NI) dominate in general, but
occasionally one needs to include the above quasi-elastic parts
(NE).

In inclusive spectra one distinguishes the following kinematic
regions :

a) Deepest Inelastic Scattering  for $x\lesssim 0.2$, with
characteristic (anti-) screening effects.

b) NI-dominated Deep Inelastic Scattering (DIS) region, $0.2
\lesssim x \lesssim x_r(Q^2)$ with $x_r(Q^2)\approx \bigg [
(M_R^2-M^2)/Q^2+1 \bigg ]^{-1}$, the Bjorken  $x$ for resonance
excitation.

c) NI-NE interference region for $x_r(Q^2) \lesssim x\lesssim
0.85-0.95$.

d) Quasi-Elastic (QE) region around the Quasi Elastic Peak (QEP),
$0.95\lesssim x\lesssim 1.05$, dominated by NE, and only weakly
perturbed by NI tails, provided $Q^2 \lesssim (4-5)$ GeV$^2$.

e) 'Deep Quasi-Elastic' (DQE) region, $x\gtrsim 1.05$, dominated by
NE. Cross sections there are very small in comparison with those in
regions a)-d) and again, are for not too high $Q^2$, only weakly
perturbed by inelastic tails.

In the following all measured total cross sections, whether
published as EMC ratios of the above in a limited kinematic range,
or as yet unpublished results for the entire measured kinematic
ranges \cite{seely,ag}, shall be referred to as 'data'.

This note is organized as follows. We first report in Section IIa on
general features of total inclusive cross sections, which we
illustrate by a few samples for iso-singlet targets. From a
comparison of experimental and computed total cross sections over
the larger part of the kinematic $x$ range $0.35\lesssim x\lesssim
0.95$, we conclude that in the DIS region NI components are
apparently reliably computed.

For increasing $x$ towards the QE region, NE components grow, start
to compete with NI, and for not too large $Q^2$, finally overtake
those. We show that around the QEP and in both  wings, theory and
data for NE applied to ND almost never agree.

Particular attention is paid to the $A=$3 iso-doublet, where we
distinguish between $p,n$ as struck nucleons (Section IIB). Section
IIC deals with EMC ratios derived from the material in Sections
IIA,B. Since in the only publication thus far, the prime interest is
a sample of EMC ratios for the lightest nuclei in the classical EMC
region $x_{min}^{data}\lesssim x \lesssim 0.9$, a comparison with
computed results, including for $^3$He, is limited to those.

In Section III we focus on the QE region and try to extract the
reduced magnetic FF $\alpha_n=G_M^n/[\mu_n G_d]$ ($G_d$). We apply a
previously formulated criterion, which has to be fulfilled before
one can attempt an extraction. For ND data it appears virtually
never fulfilled.

As an alternative source we include in Section III the CLAS data for
$F_2^D$ \cite{osipd}. Removing in the QE region a few manifestly
scattered data points, the above-mentioned criterion is
satisfactorily met by the CLAS data, which we endowed with $(2-3)\%$
systematic errors. We shall show that the extracted averaged reduced
neutron magnetic FF $\alpha_n$ agree with the OD results.

In Section IV we exploit the same CL data in order to extract the
neutron SF $\,F_2^n$ along lines used in the past \cite{rtf2n}. In
the concluding Section we discuss both theoretical and experimental
aspects of the extraction of $n$ properties from inclusive cross
sections.

\section{Cross sections and derived observables.}

\subsection {Total inclusive cross sections.}

Total inclusive cross sections are usually computed from forms like
Eq. ({\ref{a2b}), with $f^{A}(x,Q^2)$ in some approximation, for
instance the PWIA (or DWIA), or on the light cone. Below we adhere
to a non-perturbative GRS theory, which we have exploited over years
\cite{asr1,gr}. Obviously, starting from one given Hamiltonian,
different approaches evaluated to sufficiently high order in
suitable expansion coefficients, should ultimately tend to the same
final results \cite{rj}. Our choice of the GRS approach is only
motivated by, in general, better convergence of low order terms.

We start with an outline of the derivation for $f_p=f_n$, referring
for details to Refs. \cite{asr1,gr} and \cite{vkr}. In Eq. (2.8) of
the latter reference we mentioned and discussed the GRS expansion
\begin{eqnarray}
\phi(q,y_G)=\sum_n \bigg[\frac{M}{q}\bigg]^n \phi_n(y_G)
 \label{a101}
 \end{eqnarray}
of a related function $\phi(q,y_G)$ of $q=|\vec q|$, the 3-momentum
along the $z$-axis, and of a relativistic generalization of the
non-relativistic West scaling variable
\cite{gur}$\,\,(\eta=Q^2/4M^2$)
\begin{mathletters}
\begin{eqnarray}
 \nonumber
 y_G&=&y_G^{\infty}\bigg (1-\frac{1-x}{2(A-1)(1+x^2/\eta)}\bigg )\\
 \nonumber
 y_G^{\infty}&=& \frac{M} {\bigg (1+x^2/\eta \bigg)^{1/2}}(1-x)
 \label{a103}
 \end{eqnarray}
 \end{mathletters}
$y_G^{\infty}$ is that variable for an infinitely heavy recoiling
spectator. In the following we retain only the two lowest order
terms in the GRS series (\ref{a101}) (we abbreviate $y_G$ by $y$).

\begin{eqnarray}
\phi^{A;GRS}(q,y)\approx \phi^{A;(0)}(q,y)+ (M/q)\phi^{A;(1)}(q,y)
 \label{a44}
\end{eqnarray}
The lowest order term is expressed by means of the single-hole
spectral function (Spft) $S(E,k)$ of the target (Ref. \cite{vkr},
Eq. (2.9)) and the second term describes the dominant FSI.

The appearance of Spfts in expressions for the lowest order SF is
common to all approaches. Those are mainly distinguished by the
definition or choice of the 4th component of the missing 4-momentum.
For instance in the GRS theory the latter is determined by the
requirement, that the ejected $N$ and the spectator shall, to equal
measure, be off their mass-shells \cite{gur}. Its contribution
$\tilde {\phi^{(0)}}$ is detailed in Ref. \cite{gr}, Eqs. (66),(67).

The next order GRS term for the usually dominant FSI reads (cf. Ref.
\cite{vkr}, Eq. (2.17a))
\begin{eqnarray}
(M/q)\phi^{A;(1)}(y,q) = (M/q)\int_0^{\infty}\frac {ds}{2\pi}e^{isy}
 \int\int d{\vec r}_1 d{\vec r}_2 \rho_2^A({\vec r}_1,{\vec
r}_2;{\vec r}_1',{\vec r}_2)[i{\tilde \chi}^A_q({\vec b},z;s)],
 \label{a206}
\end{eqnarray}
contains two components. The first is a two-particle density matrix
$\rho_2$, not diagonal in one coordinate, which in principle may be
obtained from the product of two $A$-particle ground state wave
functions, integrating out $A-2$ coordinates. For $A>4$ one usually
makes a short-cut, using an interpolating approximation \cite{asr1}
\begin{eqnarray}
\rho_2^A({\vec r}_1,{\vec r}_2;{\vec r}_1',{\vec r}_2))\approx
\rho_1^A(r_1)\rho_1^A(r_2) \bigg [\frac {\rho_1^A(r_1,r_1')}
{\rho_1^A(r_1)}\bigg ] \sqrt{g^A(|{\vec r}_1-{\vec r}_2|)g^A(|{\vec
r}_1'-{\vec r}_2|)},
 \label{A3}
 \end{eqnarray}
where  $\rho_1^A(1,1)=\rho_1^A(1)$ and $g^A$ are the single-particle
density and the pair-distribution function. Using the
Negele-Vautherin Ansatz \cite{nv}, the non-diagonal single-particle
density $\rho_1^A{\vec r}_1,{\vec r}_1')$ is computed from
\begin{eqnarray}
\nonumber
 Y^A(s)& \equiv &\frac{\rho_1^A({\vec r}_1,{\vec r}_1')}{\rho_1^A(r_1)}
\nonumber\\
 \approx \int \frac{d^3k}{(2\pi)^3} e^{i{\vec k}{\vec s}}\,\,
n^A(k)&=&\frac{1}{2\pi^2s}\int_0^{\infty} dk \,k\,{\rm sin}(ks)
n^A(k)
 \label{a31}
 \end{eqnarray}
$n^A(k)$ is the single particle momentum distribution, obtained by
integrating the Spft over the missing energy $E$
\begin{eqnarray}
 n^A(k)=\int_0^{\infty} dE S^A(k,E)
\label{a32}
 \end{eqnarray}
The second factor in the integrand in (2.3) is an off-shell phase
factor ${\tilde \chi}_q({\vec b},z;s)$ in terms of the relative
coordinates ${\vec r}_1-{\vec r}_2=({\vec b},z)$; $\,\, {\vec
s}\equiv{\vec r}_1-{\vec r}_1'=(\vec {q}/|q|s)$. The appended
$q\approx|\vec p+\vec q|$ is approximately the lab momentum of the
nucleon, which absorbed the virtual photon, before a FSI scattering
from another $N$ occurs.

In Ref. \cite{vkr}, Eqs. (2.17b)-(2.21b) we discuss the
approximation
\begin{eqnarray}
 i{\tilde \chi}_q(\vec b,z)\approx
 \theta(z)\bigg [\theta(s-z)-s\delta(s-z)\bigg ]\Gamma_q(b),
\label{a78}
\end{eqnarray}
with $\Gamma^{(1)}_q(b)=-(\sigma_q^{tot}/2)(1-i\tau_q )A_q(b)$ the
standard on-shell profile function. It is related to the diffractive
elastic $NN$ scattering amplitude $f_q^{NN}$, with $\tau_q={\rm
Re}f_q/{\rm Im}f_q$ and $A_q(b)=[Q_q(0)]^2/4\pi]
e^{-[bQ_q^(0)]^2/4}$. With $np$ and $pp$ data of quite different
quality, one usually takes an average of the relevant $np$ and $pp$
cross sections \cite{asr1}.

Next one transforms the representative terms $\phi(q,y_G)$ in the
GRS expansion (\ref{a44}) by means of a Jacobian
\begin{eqnarray}
J^A(x,Q^2)&=&\bigg |{\partial y^A}/{\partial x} \bigg|
 \nonumber\\
 &\approx&M\bigg [\frac {1+x/\eta}{(1+x^2/\eta)^{3/2}}\bigg ] \bigg |1-
 \frac{(1-x)(2+3x/2\eta-x^2/\eta)}{2(A-1)(1+x/\eta)(1+x^2/\eta)}\bigg|
 \label{a29}
\end{eqnarray}
to the distribution function $f^A$ in the $x,Q^2$ variables
\begin{eqnarray}
\phi^A(q,y) \to f^A(x,Q^2)) = J^A(x,Q^2) \phi^A\bigg
(q(x,Q^2),y^A(x,Q^2)\bigg )
 \label{a83}
\end{eqnarray}
By means of those distribution functions $f^A$ one computes the SFs
$F_k^A(x,Q^2)$ in Eq. ({\ref{a2a}), and in particular the inelastic
parts NI$^{calc}$ \cite{asr1,rtv}. The elastic components NE are
expressed in terms of FFs and the computed distribution functions
$f^A(z,Q^2)$ as in Eqs. (\ref{a3a}), (\ref{a3b}). Their sum defines
total cross sections
\begin{eqnarray}
\sigma^{A,tot}={\rm NI}^{calc}+{\rm NE}^{FF}
 \label{a88}
\end{eqnarray}
The above we applied to all E03-102, E02-90 total cross section
data.

In view of the fact that only a restricted part of the measured ND
data have been published, we first display in Figs. 1a-d, 2a-d a
sample of $I$=0 targets, namely D($\theta=18^{\circ}, 22^{\circ},
26^{\circ},32^{\circ}$) and C ($\theta=26^{\circ}, 32^{\circ},
40^{\circ}, 50^{\circ})$. Those are shown as heavy drawn lines, to
be distinguished from heavy dots for data, first shown without error
bars.

With exception of remnants of resonance excitations of the lightest
nuclei at low $\theta$ (i.e. low $Q^2$), the examples of smooth data
shown are typical for all targets $A\ge 12$ at similar kinematics.

A cursory glance on the logarithms of the above total cross sections
shows reasonable agreement for each scattering angle and target, in
particular for the smallest $x$ measured. This holds down to the
approach of the resonance region, where NI$^{calc} \gg$ NE. The
read-off agreement thus provides evidence that the calculated NI
components in the DIS region are basically correct. In contrast,
when moving to the QE and DQE regions, growing discrepancies occur
in very small cross sections. In order to understand the nature of
the above discrepancies, we separately consider NE and NI
components.

In addition to the computed NE$^{FF}$, Eqs. (1.4})-(1.5) we define
\begin{eqnarray}
\rm{NE}^{\it extr} = \rm{data} - NI^{\it calc}
 \label{a4}
\end{eqnarray}
Clearly, the semi-empirical NE$^{extr}$ will show the scatter
present in the data, while NE$^{FF}$ in $\sigma^{A,tot}$, Eq.
(\ref{a88}), is a smooth function of $x$, $f^{A}$ and FFs. For
perfect data and theory NE from Eqs. (\ref{a3b}) and (\ref{a4})
should coincide.

The expression NE$^{FF}$ in the single photon exchange (SPE), Eq.
(\ref{a3b}), uses among others $G_E^p$. Primarily a discrepancy in
the ratio $\gamma= \mu_pG_E^p/G_M^p$, once measured in a Rosenbluth
separation, and then by polarization transfer \cite{jon}. It has led
to calculations of two-photon exchange (TPE) corrections on an
isolated $p$ \cite{blun,amt}). Those have been shown to reduce the
above mentioned discrepancy.

The above TPE have  been parameterized in the functional form of a
SPE part, which enables the sum SPE+TPE to be considered as an
effective SPE \cite{amt}. All results in the following use that
input. Their effect on single bound $p$ appears to change $\sigma^A$
for 'bare' SPE by less than $1\%$. There also exist TPE corrections
involving two nucleons, etc., but their contributions have as yet
not been determined.

We return to Figs. 1a-d, 2a-d where we display NI$^{calc}$ (light
dots) and NE$^{FF}$ (light drawn lines), as well as their calculated
sum $\sigma^{A,tot}$. Crosses in the above Figs. represent
NE$^{extr}$, Eq. (\ref{a4}) and those are seen to differ
considerably from NE$^{FF}$. Missing crosses indicate that
NI$^{calc}$ locally exceeds data.

For growing $x$, NE increasingly competes with NI and eventually
dominates, and we thus focus on the QE region. Although on a
logarithmic scale a small number of points in the QE regions may
occasionally seem to be close to the dotted lines, actual
discrepancies come to the fore on linear plots for NE$^{PP}$
together with NE$^{extr}$. The latter now include total error bars,
where we added to statistical errors (2-3)$\%$ estimates for
systematic ones. Figs. 3a-d  are samples for D(18,40), C(18,40).
Discrepancies appear to grow with both $A$ and $Q^2$ and are
occasionally quite erratic. Clearly the ND cross sections and the
results of standard computations are at odds.

This is not the case for the analysis of the OD data, where we
applied the same code. We illustrate this by a comparison with Figs.
4a,b,c for $\sigma^D(E=4.045$ GeV;$\,\theta=15^{\circ}, 30^{\circ},
55^{\circ})$, taken from Ref. \cite{rtv}. The logarithmic scale and
Bjorken value on the vertical and horizontal scales, as well as
symbols and curves correspond to those in the above Figs. 1,2. We
added dashed curves for empirical inelastic NI parts, which cause
NE$^{PP}\approx {\rm NE}^{extr}$. In the critical region between the
QEP and the (first) resonance, NI$^{emp}$ exceeds NI$^{comp}$ by
less than 15$\%$. It is obvious that in the QE region OD data and
computed results agree far better than is the case for the ND. We
shall return to this issue in the Discussion.

\subsection{The $A$=3 iso-doublet.}

Amongst the ND are also the first results for ${^3}$He after the old
SLAC data \cite{rock}. Those are of particular interest, since
$^3$He is the lightest stable odd nucleus with a large relative
nucleon excess. Approximate charge symmetry invites a simultaneous
study of $^3$He and $^3$H, although for the latter there are as yet
no data.

We thus separately treat $N=p,n$ and start with the lowest order
term $\phi^{A;(0)}$, given in I, Eq. (2.9) (or equivalently Eqs.
(66), (67) in Ref. \cite{gr}). This clearly demands knowledge of
Spfts  $S^{A=3}_N$ for the ejected nucleon $N=p,n$. In those one
distinguishes between a 2-body continuum and a D spectator state.
The latter occurs only if $I_3^{A=3}= I_3^N$ for the 3-components of
the iso-spins.

Using $\psi^{A=2}_n$ for spectator states with separation energy
${\cal E}_n^{A=2}$, one writes for the Spft  and momentum
distribution $n^{A=3}(k)=\int_{E_{min}} dE,\,\ S^{A=3}(k,E)$ of a
$A=3$ nucleus
\begin{mathletters}
\begin{eqnarray}
\nonumber
 S^{I_3^{A=3}}(k,E)&=& \sum_{j\ne D}|\langle
\Psi_0^{I_3^{A=3}}|\Psi_{j\ne D}^{I_3^{A=2}}*\vec k;I_3^N
\rangle|^2\delta(E-{\cal E}^{A=2}_{j\ne D})\\
 &+&\delta(I_3^{A=3},I_3^N)
|\langle\Psi_0^{I_3^{A=3}}|\Psi^D*\vec k;I_3^N
\rangle|^2\delta(E+B_D)\\
 \label{a107}
 n^{I_3^{A=3}}(k)&=&\sum_{j\ne D}|\langle
 \Psi_0^{I_3^{A=3}}|\Psi_{j\ne D}^{I_3^{A=2}}*\vec k;I_3^N
\rangle|^2 + \delta(I_3^{A=3},I_3^N) |\langle
\Psi_0^{I_3^{A=3}}|\Psi^D* \vec k;I_3^N\rangle|^2,
 \label{a108}
\end{eqnarray}
\end{mathletters}
with $B_D$ the binding energy of the D. One then derives the
corresponding lowest order $A=3$ distribution functions
($\beta=M\nu/q;\,\, \xi=E/M+x-1$)
\begin{mathletters}
\begin{eqnarray}
 \nonumber
 && f_N^{I_3^{A=3};(0)}(x,Q^2)= \frac {J^{(3)}(x,Q^2)}{4\pi^2}
  \bigg [\theta(x-1)\int_0^{\infty} dE \int_{\beta\xi}^{\infty} dk k S_N^{I_3^{A=3}}(k,E)
  +\theta(1-x) \\
 \nonumber
 &*& \bigg(\int_{M(1-x)}^{\infty} dE
   \int_{\beta\xi}^{\infty}dkk
     S_N^{I_3^{A=3}}(k,E) + \int_0^{M(1-x)} dE \int_{-\beta\xi}^{\infty} dk k
     S_N^{I_3^{A=3}}(k,E)\bigg )\bigg ] \\
 &+&\delta(I_3^{A=3},I_3^N)\int_{|y|}^{\infty} dkkn^{A=3}_{N;D}(k),
 \label{a20}
\end{eqnarray}
\end{mathletters}
with $y={\rm lim}_{E\to-B_D}\,\,(\beta\xi)$. $J^{(3)}(x,Q^2)$ above
is the Jacobian, Eq. (\ref{a29}) and
\begin{eqnarray}
 n_{N;D}^{A=3}(k)=\int_0^{\infty} dE S_N^{I_3^{A=3}}(k,E)\delta(E+B_D),
 \label{a26}
\end{eqnarray}
the D-component of the $A=3$ momentum distribution.

Regarding the dominant FSI term, Eq. ({2.3), one distinguishes as
before between a nucleon '1' being a $p$ or a $n$, etc. In the
evaluation the following assumptions shall be made:

i) $p,n$  number densities $\rho_1$ \cite{deV}  are equal in either
$^3$He or $^3$H, but not in both species.

ii) As to single $p,n$ momentum distributions, we computed
$n_{p,n}^{^3{He}}(k)$ from the generalization Eq. (\ref{a108}) of
(\ref{A3}) and found small differences for single $p$ and $n$
components.

iii) In the generalization of Eq. (\ref{A3}) we assume
$g^{nn}(|{\vec r}_1-{\vec r}_2|)=g^{pp}(|{\vec r}_1-{\vec r}_2|)$
and $g^{pn}(|{\vec r}_1-{\vec r}_2|)=g^{np}(|{\vec r}_1-{\vec
r}_2|)$, which change Eq. (\ref{a206}) into
\begin{mathletters}
\begin{eqnarray}
 \nonumber
&&\rho_2^{I_3^{A=3}}(1,2;1'2)\to \\
\nonumber
 & 1/3&\bigg [
 \rho_2({\vec r}_{1,p},{\vec r}_{2,p};{\vec r}'_{1;p},{\vec r}_{2,p})+
 \rho_2({\vec r}_{1,p},{\vec r}_{2,n};{\vec r}'_{1,p},{\vec r}_{2,n})+
 \rho_2({\vec r}_{1,n},{\vec r}_{2,p};{\vec r}'_{1,n},{\vec r}_{2,p})
 \bigg ]_{I_3^{A=3}}\\
\nonumber\\
 &\approx &1/3 \bigg [Y(s)\rho_1({\vec r}_1) \rho_1({\vec r}_2)\bigg ]_{I_3^{A=3}}
 \bigg [2\sqrt{g_{pn}(r)g_{pn}(|{\vec r}-{\vec s}|)} +
         \sqrt{g_{pp}(r)g^{pp}(|{\vec r}-{\vec s}|)}\bigg ]
 \label{a120}
\end{eqnarray}
\end{mathletters}
The functions $g_{pp}$ and $g_{pn}$ have been taken from Ref.
\cite{sss}. Neglecting the small differences in the single $N$
momentum distributions $n_N^{I_3^{A=3}}$, the points i)-iii) leave
no $N$ dependence: $\phi^{A;(1)}$ depends only on $I_3^{A=3}$.

Next we replace ${\vec r}_1,{\vec r}_2$ by relative and CMS
coordinates ${\vec r},{\vec R}$ and perform the $R$-integration in
(\ref{a206})
\begin{eqnarray}
T({\vec b},z)=\int d^3R \rho_1(|{\vec R}+{\vec r/2}|)\rho_1(|{\vec
R}-{\vec r}/2|),
  \label{a832}
\end{eqnarray}
which leaves one angular and one radial integration in the
expression for ${\tilde \phi}^{(1)}$. Consequently, $Y$ in Eq.
(\ref{a120}) appears as a factor in
\begin{eqnarray}
\int d{\vec R}\rho_2^A(1,2;1',2) \to \frac{1}{3}Y^A(s)T^A(b,z) {\cal
G}({\vec b},z;s),
  \label{a60}
\end{eqnarray}
with ${\cal G}$ a combination of pair-distribution functions in
({\ref{a120}).

Finally, using approximation (2.7) for the phase, the FSI
contribution (2.3) becomes
\begin{mathletters}
\begin{eqnarray}
\nonumber
 \phi^{A=3;(1)}(q,y)&=&\frac{1}{3}{\cal R}e\int_0^{\infty} ds{\rm e}^{isy} Y(s)\int_0^{\infty}db
 b\\
 \nonumber\\
 &&\bigg [\int_0^s dz \bigg ( T(b,z){\cal G}(b,z;s)\bigg )-sT(b,s) {\cal G}(b,s;s)\bigg],
 \label{a61}
\end{eqnarray}
\end{mathletters}
where 6-dimensional integrals in Eq. (\ref{a206}) are reduced to
2-dimensional ones. Again, Eq. (2.6) in Ref. \cite{vkr} produces the
corresponding $f^{A=3;(1)}(x,Q^2)$. We checked that $f^{FSI;(1)}\ll
f^{(0)}$, but did retain the FSI term $f^{(1)}$ in all calculations.

We now reach the crucial input, which is the outcome of extensive
calculations for various Spfts  $S^{{^{3}He}}$, performed by the
Rome-Pisa group \cite{gianni}. Those employed the following $NN$
interactions:

a) purely 2-body $NN$ forces (B2; AV18 \cite{wir}), neglecting
$V_{coul}$.

b) the same as a), including $V_{coul}$.

c) the same as b) with an additional 3$N$ force (B2+B3; AV18 UR9
\cite{pud}).

The list above does not refer to $^3$H. All items were intended as
input for $^3$He calculations but clearly, only b) and c) are
realistic options with different sophistication. In contrast, option
a) lacks $V_{coul}$ between protons. i.e. the most obvious and
dominant charge-symmetry breaking part and is therefore not suited
for $^3$H3 calculation.

However in the absence of other parts, the missing $V_{coul}$ turns
the hamiltonian for a) to the the charge-symmetric one corresponding
to b), i.e. for $^3$H.

However, option a) is for a $^3$He Hamiltonian which lacks Coulomb
forces between the protons. Thus disregarding additional
charge-symmetry breaking effect, option a) describes the iso-partner
$^3$H. In particular for the basic distribution functions one has
the following relations for the iso-partners
\begin{eqnarray}
f_n^{{^3}H}=f_p^{{^3}He(no\,\,coul)} ;
f_p^{{^3}H}=f_n^{{^3}He(no\,\, coul)}
\label{a50}
\end{eqnarray}
Using Eq. (\ref{a2b}) one has for the SFs
\begin{eqnarray}
F_2^{{^3}He}(x,Q^2) = \int_x^3 dz\frac{1}{3}
\bigg[2f_p^{{^3}He}(z,Q^2)F_2^p\bigg(\frac{x}{z},Q^2\bigg)+f_n^{{^3}He}(z,Q^2)F_2^n\bigg(\frac{x}{z},Q^2\bigg)
\bigg]
 \label{a49}
\end{eqnarray}
and either form
\begin{mathletters}
\begin{eqnarray}
 && F_2^{{^3}H}(x,Q^2)=\int_x^3 dz\frac{1}{3}
\bigg[f_p^{{^3}H}(z,Q^2)F_2^p\bigg(\frac{x}{z},Q^2\bigg)+2f_n^{{^3}H}(z,Q^2)F_2^n\bigg(\frac{x}{z},Q^2\bigg)
\bigg ]
 \label{a57}\\
 &=& \int_x^3 dz\frac{1}{3}
\bigg[f_n^{{^3}He(no\,\,coul)}(z,Q^2)F_2^p\bigg(\frac{x}{z},Q^2\bigg)+2f_p^{{^3}He(no\,\,coul)}(z,Q^2)
F_2^n\bigg(\frac{x}{z},Q^2\bigg)\bigg ]
 \label{a109}
\end{eqnarray}
\end{mathletters}
Next we compare the above considerations with some results. Figs.
5a,b show for $Q^2=2.5$, respectively 7.5 GeV$^2$ the distribution
functions $f_{p,n}^{{^3}He}(x,Q^2)$, Eq. (\ref{a29}). The drawn,
dashed and dotted correspond to interactions B2(0), B2(0+1),
B2+B3(0+1), where the numbers indicate the order of terms retained.
Since FSI terms are retained, results for B2(0) only serve to
indicate the relative importance of the two terms. Fig. 5c compares
$f_p^{A=3,I_3}(x,Q^2=5)$, Eq. (\ref{a50}). to which we shall return
in Section IIC.

At this point we need and mention, that for any $Q^2$ the norm of
the lowest order term ${\cal N}_x=\int_0^3
dxf_{p,n}^{A=3;(0)}(x,Q^2)=1$. A more extensive discussion can be
found in the Appendix.

The above $A=3$ distribution functions follow a standard pattern for
all light $A$ \cite{rtvemc}: for increasing $Q^2$ the peak of
$f^{A=3}$ increases and the width shrinks correspondingly. For
instance, for $Q^2$ increasing from 2.5 to 10.0 GeV$^2$, the B2 $p$
peaks increase from 4.041 to 5.394  and for $n$ from 3.483 to 4.648.
Peak values for B2+B3 are $\approx 4\%$ lower than for B2 and are
correspondingly wider. Next, for the same $Q^2$ those are
intermediate between the same for D and $^4$He.

Fig. 6 displays computed SFs $F_2^{{^3}He}$ for fixed
$\theta=40^{\circ}$ and variable $Q^2(x,\theta)$, using Eqs.
(\ref{a49}) with B2 or B2+B3 interactions. The results for those are
practically indistinguishable and quite close to the extracted SF,
using data and $R\approx 0.36/Q^2$ for the ratio of inclusive
scattering of virtual longitudinal and transverse photons. The lower
curve in Fig. 6 is for $F_2^{{^3}H}$, Eq, (\ref{a109}): with no
data, there is no extracted parallel. One notices however, the
seizable difference in predictions for the members of the
iso-doublet.

Figs. 7a,b present on a linear scale data and computed inclusive
cross sections on $^3$He for $\theta=18^{\circ}, 40^{\circ}$. Cross
sections for $\theta=40^{\circ}$ for $^3$He, using either B2 or
B2+B3 interactions are in good agreement with data. This outcome
should be compared with the same for other targets shown in Figs. 1
on a log scale, and the same for the QE range in Figs. 2 on a linear
scale. The rather poor fit for $^3$He, $\theta=18^{\circ}$ is in
striking contrast with the same for $\theta=40^{\circ}$, in spite of
the use of the same underlying analysis. We cannot forward any
theoretical explanation.

\subsection{EMC ratios.}

At least some difficulties in understanding ND total cross section
data may be due to unknown systematic errors, the size of which can
only be estimated. Since the D is amongst the targets, part of those
errors may cancel in EMC ratios $\mu^A=F_2^A/F_2^D$. For that reason
alone is it of interest to compare measured with computed ratios.

The new EMC data are for a few discrete $\theta$ and thus not for
fixed $Q^2$. Although the resulting $Q^2$ dependence is mild, it
should be borne in mind that, for instance, for the chosen angle
$\theta=40^{\circ}$, data on the released, or measured additional
$x$-ranges, cover $2.80\lesssim Q^2 {\rm (GeV}^2)\lesssim 6.12$,
which is not an insignificant variation in $\mu(Q^2)$. The published
data are for $x_{min}^{data} \leq x \lesssim 0.9$,
$x_{min}^{data}\approx 0.35$ \cite{seely}, and  correspond to what
occasionally is called the 'classical' EMC range. Within that range
the variation of $Q^2$ has less spread and causes less than $1\%$
variations in EMC ratios.

Measured $\mu^{^{4}He}$ and $\mu^{^{12}C}$  \cite{seely} have a
somewhat smaller slope than older data, in particular for $^4$He
(see for instance Ref. \cite{kp}. In Figs. 8a,b we compare the new
data with previously computed GRS results for $Q^2=3.5, 5.0$ GeV$^2$
\cite{rtvemc} and for additional $Q^2$, close to the above-mentioned
binned ones. The agreement is reasonable.

We also mention a prediction, based on the $Q^2$-independence of
$F_2^{p,D}(x\approx 0.20,Q^2)$ \cite{arneo}. Since all distribution
functions $f^A$ are negligible for $x\lesssim 0.4$, one may replace
the lower integration limit in the expression (2.1) for $F_2^A$ by
0. Then using unitarity, i.e. ${\cal N}[f^A]\approx {\cal
N}[f^{A;(0)}]=1$, all $F_2^A(x\approx 0.20,Q^2)$ computed by Eq.
(\ref{a2a}) are predicted to be roughly independent of $Q^2$ as well
as of $A$. Consequently EMC ratios $\mu^A(x,Q^2)$ ought to intercept
the $x$-axis at a value $\mu^A(x\approx 0.2,Q^2)\approx1$
\cite{rtvemc}. The above holds when only one distribution function
is involved, i.e. for $I=0$ nuclei, or when an averaged $f^A$ is
sufficiently accurate in all other cases.

The actual crossover $\mu^{A,D}=1$ for most nuclei and for several
$\mu^{A,A'}$) is $x_{co}\approx (0.2-0.3)$, whereas the intercept of
the $I=0$ ND data, seems to occur for somewhat higher $x\approx
0.33$ \cite{seely}.

Again we separately discuss the case of the $A=3$ iso-doublet with
different $f_p$ and $f_n$. In the previous Section we presented
results for computed SFs and cross sections for the $A=3$ doublet,
using various input options. For completeness we add for $^3$He the
'standard' extraction of its SF from cross section data. Although
the above exists for several options, we report only on computed
results for B2+B3 interactions and calculated EMC ratios as
$F_2^A/F_2^D$, respectively $\sigma^A/\sigma^D$.

As discussed above we need option a) in Section IIB for $^3$He with
no $V_{coul}$, in order to compute the the SF for iso-partner $^3$H.
Data are much desired \cite{petra} \cite{pace}, but it will take
years before those will become available and can be confronted with
calculated $F_k^{^3{H}}$, like from Eq. (\ref{a109}) (cf. also
\cite{sss}, \cite{pace}, \cite{afnan}).

In Fig. 9a we show 3 curves for $\mu^{A=3;F_2}(\theta=40^{\circ})$
from ratios of SFs. The upper and lower ones are for $^3$He and
$^3$H, while the middle one for half their sum, is the computed
$I=0$ part of either member of the iso-doublet. One notices the
widely different behavior of the two ratios: In the classical EMC
region $\mu^{^3{He}}>1,$ has a positive $x$-slope and shows no
minimum for medium $x$. In contrast $\mu^{^3{H}}<1$, has a negative
$x$-slope and an unexpectedly deep minimum for $x\approx 0.7$. The
iso-singlet part has positive slope, crosses 1 at $x\approx 0.8$ and
has no visible minimum.

Fig. 9b shows $\mu^{A=3;\sigma}(\theta=40^{\circ})$, but now as
ratios of cross sections, which are seen to differ from the one in
Fig. 9a: $\mu^{^{3}He}\gtrsim 1$ for $x\lesssim 0.85$ and has a
maximum for $x\approx 0.75$. In contrast $\mu^{^{3}H}<1$, has
negative $x$-slope and shows a shallow minimum for $x\approx
(0.6-0.7)$. Essentially the same holds for the iso-scalar part, but
the $^3$He part there pushes the $I=0$ part an amount $\approx 0.1$
upwards on the $\mu$-scale.

The empty circles in Fig. 9b are the data of Seely $et\, al.$ for
$^3$He from ratios of bona fide cross sections, yet are called by
the authors 'raw data' (empty circles in Fig. 9b). Comparison with
the upper drawn curve shows rough agreement. In contrast to a
genuine calculation of the iso-scalar part (dashed curve), the
above-mentioned authors modify the above 'raw data' in a standard
fashion, which does not require information on $^3$H. This leads to
a fictitious $I=0$ nucleus with $N=Z=A/2$ amounting to
\begin{eqnarray}
f_{p,n}^{A(Z,N)} \approx f^A(Z=N)\approx
f^A\bigg(\frac{A}{2},\frac{A}{2}\bigg)
 \label{a501}
\end{eqnarray}
The above is considered to be a model for the EMC ratio of an even
nucleus with $I\ne0$ and instructive, even for interpolation to
$A=3$ \cite{seely}.

The above results hardly change when $\theta$ runs over the entire
measured range, and this holds in particular for their $I_3$
dependence. One should keep in mind that the anyhow small EMC effect
is the deviation from 1 of the ratio of small numbers. That effect
even diminishes when going to the lightest nuclei, increasing its
precarious sensitivity. In this relation we recall the $I$-spin
dependence in Fig. 5c of the $p$-distribution function in the
iso-doublet, assuming iso-spin symmetry after correcting for
$V_{coul}$. The two may well be related.

\section {The magnetic FF of the neutron and CLAS data.}

In the previous sections we encountered clear discrepancies between
the NE component in the ND and OD results in the QE regions of total
inclusive cross sections. Their description requires the dominant
reduced $n$ magnetic FF $\alpha_n=G_n^M/[\mu_nG_D]$ in the QE
region. In the following we recall attempts to isolate and to
extract $\alpha_n$.

At this point we remark that on the one hand the entities $G_M^n$
(and $F_k^n$ considered in the following section) are needed to
determine the total SFs $F_A$ and the inclusive (reduced) cross
sections $\sigma^A$. On the other hand one wishes to extract those
from QE data. The procedure is to use some starting values in the
input, compare the output until self-consistency is reached and the
outcome compared with the staring values.

The expressions (\ref{a3a}), (\ref{a3b}) locate two functions with
pronounced peaks for $x\approx 1$. Those are the NE part of the
reduced cross section $K^{A:NE}(x,Q^2;E,x)$, Eqs. (1.1), (2.11), and
the linking distribution function $f^A(x,Q^2)$, functions of 5,
respectively 3 variables.

In appears that their ratio $K^{A;NE}/f^A$ is primarily a function
of $Q^2$ with only weak additional dependence on $x, \theta, E$ and
$A$. As suggested in the past we turn the above into a criterion, to
be fulfilled by candidates $x_l$ for extraction

For sufficiently accurate and smooth data one tries to locate a
$continuous$ $x$-range in the QE region, for which the above ratio
does not vary, by more than a prescribed amount, say, 10$\%$. If
available, one finds (cf. Eq. (4.3) in Ref. \cite{rtv})
\begin{eqnarray}
 \alpha_n|\mu_n|=\bigg[\frac{2MK^{A,NE}/[vfG_d]-B^2/\eta}
{1+{\rm tan}^2(\theta/2)/v}- \bigg (\alpha_p \mu_p\bigg )^2\bigg
]^{1/2},
 \label{a13}
\end{eqnarray}
where $Q^2$-dependence is implicit. Above $G_d$ is the standard
dipole FF, $v=x^2/2(1+\eta)$, $\beta_N=G_E^N/G_d\,\,\, \alpha_N=
G_M^N/[\mu_NG_d]$ and $B^2=\beta_p^2+\beta_n^2$. Eq. (\ref{a13})
generalizes for $f_n\ne f_p$ as
\begin{eqnarray}
\frac {N}{A}f_n[\alpha_n\mu_n]^2 +
   \frac {Z}{A}f_p[\alpha_p\mu_p]^2
=\frac{MK^{A;NE}/vG_d^2-\bigg( Zf_p\beta_p^2 /(A\eta)+
Nf_n\beta_n^2\bigg)}
     {1+{\rm tan}^2(\theta/2)/v},
\label{a104}
\end{eqnarray}
$\alpha_n(Q^2)$ is of course only a function of $Q^2$, but due to
imperfect data and theory the algorithm produces an inherent, weak
dependence on the chosen points $x$. Whereas there is no physical
meaning to individual $x$-dependent results, it is natural to define
the extracted $\alpha_n(Q^2)=\equiv \langle\alpha_n(Q^2)\rangle=
\langle\alpha_n(x;Q^2)\rangle_x$ as an appropriate average over the
selected $x-$range. For all previously investigated OD data the
above criterion is met for a suitable number of continuous
$x$-points (see Table in Ref. \cite{rtv}). As to ND, only for
D$(\theta\lesssim 32^{\circ})$ could we find 2-3 such points.
However, those points appear to produce through Eq. (\ref{a13}) a
value for $\alpha_n$, far from the OD results for similar $Q^2$.

Like the material discussed in Sections I, II, also the above
indicates that in the QE region the ND and OD data sets do not
match. We emphasize two points, relevant for OD. One is the very
applicability of the suggested analysis for OD data, in contrast to
the same for ND. Moreover $different$ sets with approximately the
same $Q^2$, produce essentially the same $ \alpha_n(Q^2)$, providing
evidence for internal consistency \cite{rtv}.

As a last resource we invoke the CLAS collaboration data on
$F_2^{D}$, which have not been subjected to a similar analysis
before. Those are available for a dense net of $Q^2$ ($\Delta Q^2$ =
0.05 GeV$^2$), which for each $Q^2$ cover a wide and dense $x$-range
($\Delta x$ =0.009) \cite{osipd}. We apply the above mentioned
criterion regarding the $K/f$ ratio to those data and look for
continuous $x$ ranges around the QEP for $\theta= 18^{\circ},
22^{\circ},26^{\circ}, 32^{\circ}$, which approximately correspond
to $Q^2\approx 2.50, 3.275, 4.175, 5.175 $ GeV$^2$. Regrettably CL
data do not extend to larger $Q^2$, covering $\theta= 40^{\circ},
50^{\circ}$ in ND.

Also for CL one cannot, strictly speaking, apply the above criterion
for a continuous $x$-range in every data set. However, a
representative number of candidate $x$-points remains after removal
of at most 1 or 2 points per  set, for which the observed scatter of
neighboring  points exceeds 10$\%$. The extracted
$\langle\alpha_n(Q^2)\rangle$ appears to match the OD results.

We first show in Figs. 10a,b in much the same way as in Figs. 1-2,
the components NI$^{D,calc}$, NE$^{D,FF}$ and NE$^{D,extr}$ for two
of the above four data sets with $Q^2$= 3.275, 4.175 GeV$^2$.
Whereas clearly useful around the QEP, some disagreement between the
two NE representations grows towards the inelastic wing of the QE
peak. It is similar in size and shape as for OD data \cite{rtv}, but
not anywhere as disastrous as for the above-mentioned ND.

In Table I we entered $F_2^D$ for the above four angles over a range
of $x$ and correspondingly varying $Q^2$ values. In the last 3
columns we compare: i) the values $extracted$ from the ND data,
assuming the standard transverse/longitudinal ratio $R\approx
0.36/Q^2$; ii) the same $computed$ from Eq. (\ref{a2b}); iii) the CL
data for $F_2^D$. Differences seem largest around the QEP and
occasionally switch sign. No similarly large abberations are
apparent in the analysis of linear plots of the OD data.

Table II contains the reduced magnetic FF $\alpha_n^{CL}$ from the
CLAS data, $Q^2$, the range and number of chosen $x$-points. Column
4 states the averaged $\langle \alpha_n(Q^2)\rangle$ with the error
of the mean. To the statistical errors we included guessed 2$\%$
systematic ones, added in quadrature.

In Fig. 11 we assemble $\alpha_n(Q^2)$ as extracted from the OD and
CL data for 4 values $Q^2$= 2.501, 3.275, 4.175, 5.175 GeV$^2$,
together with a previously extracted parametrization, Eq. (5.4),
Ref. \cite{rtv}. For completeness we added to the above all
$\alpha_n$ with $Q^2\ge 2.5$ GeV$^2$ \cite{rtv}, extracted  from OD
data. The CL and OD data sets produce essentially the same results
and trend.

The above figure displays $\langle \alpha_n(Q^2)\rangle$, extracted
from CL for closely spaced $Q^2$ around the 4 values above. While
the former vary by $(0.5-1.0)\%$, going from one to a neighboring
$Q^2$-bin, entries within each bin show  larger variations within a
standard deviation of $\langle \alpha_n(Q^2)\rangle$.

In the same figure we entered $\alpha_n(Q^2)$, recently extracted
from the cross section ratio $D(e,e'n)p/D(e,e'p)n$ for $Q^2=
(1.0-4.8)$ GeV$^2$. Each FF  point has been measured with a less
than 3$\%$ error, but again, for $Q^2\gtrsim 2.5$ GeV$^2$, and most
outspoken for $Q^2\gtrsim$ 3.4 GeV$^2$, the scatter between adjacent
points, is often far larger. The results (Fig. 5 in Ref.
\cite{brooks}) have also been included in Fig. 11 together with
those of Ref. \cite{lung}.

We conclude this section, mentioning a recent calculation of space
and time-like  nucleon FFs using a light-front framework
\cite{melo}. Whereas space-like $p$ FFs are well reproduced, the
computed $\alpha_n$ show a maximum, then diminishes and tends to 0
for large $Q^2$. That result disagrees considerably with those
displayed in Fig. 11, where practically coinciding extractions from
the CLAS collaboration data and the OD, produce a  continuous
decrease, which persists out to $Q^2\approx 10\,$GeV$^2$.

\section {Extraction of $F_2^n$ from CLAS data.}

The neutron SF $F_2^n$ complements information from $F_2^p$ on the
valence quark distribution functions $u_v,d_v$ in the $N$: its
knowledge is a minimal requirement to disentangle the two
distributions. Lacking reliable information on $F_2^n$, one
occasionally invoked the SU(6) result $[F_2^n/F_2^p]_{SU6}=3/5$
which amounts to $u_v=2d_v$. Alternatively, one uses the 'primitive'
choice $F_2^n=2F_2^D-F_2^p$, the reliability of which is restricted
to $x\lesssim 0.35$.

It is clearly desirable to have empirical information on $F_2^n$
along with $F_2^{p,D}$ for $x\gtrsim 0.3$ in order to determine $N$
parton distributions. With no free $n$ target available, one has to
extract $F_2^n$ from bound neutrons. The preferred target has been
the D, for which the nuclear information is simplest and most
accurately known, but the literature also describes the extraction
of $F_2^n$ from future precision data on the SFs for $^3$He,$^3$H
(e.g. Refs. \cite{pace,sss,afnan}) and heavier targets \cite{rtf2n}.

Several extraction methods have in the past been proposed. For
instance, an approximate inversion of Eq. (\ref{a2a}) requires
reliable data points on $F_2^A$, and preferably for several targets.
Previously we had found that data were barely sufficient to extract
$F_2^n$ from a single binned ${\bar Q}^2= (3.5-4.0)$ GeV$^2$. We
summarize the steps of the followed procedure \cite{rtf2n}, which
will also be exploited below:

i) Assume $C(x,Q^2)\equiv F_k^n(x,Q^2)/F_k^p(x,Q^2)$ to be
independent of $k$.

ii) $C(0)=1$, as implied by a finite Gottfried sum $\int_0^1 dx
[F_2^p(x,Q)2-F_2^n(x,Q^2)]/x$.

iii) The validity for $x\lesssim 0.30-0.35$ of the 'primitive'
approximation $F_2^D=[F_2^p+F_2^n]/2$, $i.e.$ $C=2F_2^D/F_2^p -1$.
In practice we use $p,D$ data for 2 points, chosen to be $x=0.15,
0.25$. The information ii), iii) mainly determines the decrease of
$C(x,Q^2)$ from 1 for $x$, increasing from 0 to about $x\lesssim
0.6$.

iv) A last step is a chosen parametrization for $C(x,Q^2)=\sum_{k
\ge 0}d_k(Q^2) (1-x)^k$. For $k=3\,\,$, ii)-iii) leave one parameter
to be determined and the natural candidate is
$C(1,Q^2)=d_0(Q^2)=1-\sum_{k \ge 1}d_k(Q^2)$.

A remark on $C(1)$ is in order here. Both SFs $F_2^{p,n}(x,Q^2)$
vanish for finite $Q^2$ beyond the lowest inelastic pion production
threshold at $x_{\pi\,thr}(Q^2)\approx 1/\big
[2M\mu_{\pi}/{Q^2}+1\big]$, and the NI continuum is therefore
isolated from the elastic peak at $x=1.$ The SFs of the latter are
given by Eqs. ($\ref{a3a}),(\ref{a3b}$) in terms of FFs. Neglecting
$G_E^n$, one finds (typos in Ref. \cite{rtf2n} have been corrected
below)
\begin{eqnarray}
\lim_{x\to 1}C^{FF}(x,Q^2)=
\Big[\frac{\mu_n\alpha_n(Q^2)}{\mu_p\alpha_p(Q^2)}\Big]^2\Big[1+\frac{4M^2}{Q^2}
\Big(\frac {\gamma(Q^2)}{\mu_p}\Big)^2\Big]^{-1}
\end{eqnarray}
with
\begin{mathletters}
\begin{eqnarray}
\gamma(Q^2)&=&\frac{\mu_pG_E^p(Q^2)}{G_M^p(Q^2)}
\nonumber\\
\frac{\alpha_n(Q^2)}{\alpha_p(Q^2)}&=&\frac{G_M^n(Q^2)/\mu_n}
{G_M^p(Q^2)/\mu_p}
 \label{a55}
\end{eqnarray}
\end{mathletters}
$C$ has then been determined by a least square fit for the sum
$\sum_{x_m}^{x_M}$ and not point-by-point in $x$. Naturally, any
parametrization of $C$, and in particular iv) above, ascribes values
to $C$ in the un-physical region $1\gtrsim x\gtrsim x_r(Q^2)$.

The extracted parameter $d_0(Q^2)=C(1,Q^2)$ fairly rapidly reaches a
plateau for increasing $Q^2$. Since $\lim_{x\to
1}F_2^{p,n;NI}(x,Q^2)=0$, it is not surprising to find that values
for $C(x\lesssim 1,Q^2)$ on the plateau, i.e. the extrapolation from
the adjacent non-physical region to the largest $x> x_{thr}$ and
ultimately to $x=1$, depend sensitively on the upper limit taken in
the $x$-sum above. Thus for $x_M$=0.75 and increasing $Q^2,
4\lesssim Q^2(\rm GeV^2)\lesssim 10$, the extracted $C(1,Q^2)$
decreases from 0.38 to 0.27, while $C^{FF}(1,Q^2)$, Eq. (\ref{a3b}),
barely decreases form 0.38 to 0.37. For a slightly larger
$x_M=0.80$, $C(1,Q^2)$ decreases from 0.34 to 0.25 over a much
narrower $Q^2$-interval than for $x_M=0.75$.

The procedure has been checked by a re-calculation of $F_2^D$, using
the extracted $F_2^n$ in Eq. (\ref{a2a}): the initial $F_2^D$
appears accurately reproduced. Figs. 12a,b show $C^D(x,Q^2)$ as well
as $F_2^{p,n}$ for $Q^2=2.5, 7.5$ GeV$^2$, $x_M=0.75$. Although
influencing $C$ for $x$=1, one can only barely distinguish between
$C(x\lesssim 0.85,Q^2)$, computed for either $x_M=0.75$ or 0.80.

Alternative attempts have been made in the past in order to obtain
$F_2^n$, all of which use a D target. One for instance replaces the
distribution function $f^{PN,D}$ in Eq. (\ref{a2a}) by a momentum
distribution or some generalization of the latter, and uses it to
"smear" nucleon SFs \cite{MT}. From he difference $\langle
F_2^n\rangle_f=F_2^D-\langle F_2^p \rangle_f$, featuring folded or
'smeared' SFs, the "bare" $F_2^n$ has to be de-convoluted. An
iteration method has recently been tested on the MAID
parametrization for $F_2^N$ \cite{kmk}. The reported success may in
part be due to the fact that the procedure, as well as the
parameterized input, imply the use of a smooth average. Application
to real data with non-negligible may well run into the above
discussed difficulties. We also mention an extraction of $F_2^n$
from essentially the Impulse Approximation for $f^{PN,D}$, using the
parameterized ratio $F_2^D/F_2^p$, $F_2^p$ and the D wave function
\cite{achl}. Most published $C(x,Q^2)$ follow the same trend for
$x\lesssim 0.75$ and are primarily distinguished by the extrapolated
$C(x=1)$.

Finally we recall the extraction of the leading twist moments of
$F_2^{p,D}$ from the CL data. By means of a convolution such as is
Eq. (\ref{a2a}), those are subsequently used to construct parallel
twist moments for $n$ \cite{osip3}. Also the above analysis uses
some averaging, which may in part underly its feasibility. No
inversion leading to $F_2^n(x,Q^2)$ has been attempted there.

\section{ Discussion and Conclusions.}

It has been our goal to describe the Jlab experiments E103-102,
E02-90 on inclusive electron scattering from various targets,
specifically for total cross sections and EMC ratios \cite{seely}.
Subsequently  we tried to extract from those the (reduced) magnetic
FF $\alpha_n(Q^2)$ and the SF $F_2^n(x,Q^2)$ of a neutron bound in a
nucleus. Those are respectively, the dominant part of the NE
component in the QE region and a vital component of the inelastic
part of the total cross section.

Also for the ND, we used the GRS approach which has previously been
applied  to all older experimental information \cite{rtv}. Only
minor changes in theoretical elements have been applied since, for
instance the inclusion of two-photon exchange corrections to the
electric FF of the proton.

We first mention that the most reliable results from the OD data
have been obtained in the DIS region, where inclusive scattering is
entirely inelastic. For the smallest $x$ we found agreement with
data, not rarely to within (2-3)$\%$. For increasing $x$, strongly
decreasing NI parts have to be accurately known in order to isolate
with precision the NE parts, dominating the QE region \cite{rtv}.
For medium $x$ between the "elastic tails" of (pseudo-)resonances
and their peaks disagreements appear, which are reflected in the
difference between the extracted and computed NE components.

A possible cause  of the above disagreements could be uncertainties
in the proton SF. We checked that a mild relative change of NI,
which grows to $\approx 15\%$ in the NE/NI interference region, and
again decreases towards the higher NI resonances, brings about
agreement. Such an uncertainty in the parametrization of the of
$F_2^p$ in the required $Q^2$ region, apparently hardly affects the
quality of the extracted $F_2^p$ \cite{chr1}.

Next we considered the extraction of the $n$ magnetic FF from data
in the QE region. We utilized a previously formulated criterion for
such an extraction, which requires a continuous set of eligible
$x$-points, for which the ratio of the $x$-dependent reduced total
cross section and the computed distribution function falls within
pre-determined limits. The criterion could be fulfilled for all old
data sets, which moreover showed consistency: The same $\alpha_n$
resulted from different data sets with overlapping $Q^2$ values.

From the above new precision data one expects agreement of at least
the same quality. Using exactly the same program as before we
analyzed all measured data, of which only a fraction has been
published. The major results are as follows:

a) Even in the DIS region, the best agreement is not better than
(5-6)$\%$, and not infrequently of both signs.

b) Measured EMC ratios for light iso-scalar nuclei approximately
agree with previous data and calculations .

c) The same seems to hold for model-independent features for
$x\approx 0.20$

d) It is virtually impossible to satisfy in any QE region our
criterion on candidate $x$-points  for the extraction of $\alpha_n$.
As a rule NE parts, as the difference of and computed NI components,
do not anywhere match NE, computed from FFs. The required changesin
NI leading to a match, by far exceed the moderate ones described in
Ref. \cite{rtv}.

At this point we mention a suggestion  to integrate the QE peak over
some $x$-interval and to extract $\alpha_n$ from those \cite{arrpc}.
We doubt whether the suggested procedure can produce reliable
averages for locally varying relative systematic errors.

The only alternative material which we could use in the above
analysis, are the CLAS collaboration data on D \cite{osipd}, which
have apparently not been analyzed before. From those we could
extract both $\alpha_n$ and $F_2^n$ and the former essentially
matched older results.

Particular attention has been paid to the $A=3$ iso-doublet. Many
years after the first data were taken, the new experiments contain
information on $^3$He, while also theory has much advanced. Most
significantly there are now available results of exact calculations
of the $A=3$ single $p,n$ Spectral Functions for several $NN$
interactions. Those underly the calculation of the dominant
contribution of the separate $p,n$ distribution functions in both
$^3$He and $^3$H, with the latter using charge symmetry when
$V_{coul}$ is neglected. We thus calculated for the GRS theory the
SFs of $^3$He and $^3$H and inclusive cross sections.

We start with $\sigma^{{^3}He}$ for all 6 measured angles and found
only crude agreement for the lower angles, and (very) good
correspondence for the largest ones, in particular for
$\theta=40^{\circ}$. Using one and the same theory for all, theory
cannot be blamed for the striking dissimilarity.

Next we computed the two EMC ratios $\mu^{A=3}(\theta=40^{\circ})$
and their iso-scalar mean, once as ratios of $F_2$ and then
alternatively from cross sections. The data for $^3$He in the
classical EMC region hover around 1 and do not show a minimum around
$x\approx 0.5-0.6$. About the same is predicted by theory.

The latter is quite different for $^3$H, for which theory predicts a
more standard behavior with $\mu\lesssim$ 1 and a shallow minimum.
Lack of data prevent a comparison with the above outcome. However,
one may discuss the computed iso-scalar part, which resembles a
standard EMC ratio with a growing negative slope for decreasing $A$
from $A=12$, down to D. The slope of the iso-vector part
$\mu^{A=3;I=0}(\theta=40^{\circ})$ lies in between the same for
$^4$He and the D.

As to 'raw data', in a standard procedure one estimates $\mu^A$ for
nuclei with a nucleon excess, replacing that EMC ratio by one for a
fictitious isobar with $Z=N=A/2$. The published data for the
iso-scalar part of $^3$He about agree with the computed ones. It
should be clear that theory computes a real result, while the above
data relate to a somewhat dubious extrapolation to the lowest $I\ne
0$ nucleus.

We return to the enigmatic outcome for several non-isolated ND data.
Since the same tools were used before, the most extreme conclusion
could be incompatibility of the old and new sets. A milder judgement
blames systematic errors. We had included those as a fixed estimated
percentage, but it is clear from the scatter of neighboring accurate
points, that more than average systematic errors are required in
order to bring about agreement.

We also wish to recall that all cross section data sets are reported
to have normalization uncertainties running from (2.2-2.7)$\%$
\cite{ag}. Those may in part cause some of the observed
discrepancies, but for instance not the discrepancies between
inclusive cross sections on $^3$He for $\theta=18^{\circ}$ and
$40^{\circ}$ in similar data sets.

A question of different nature is, whether a fundamental parton
description may significantly modify results based on a used
hadronic representation. Only recently has attention been re-drawn
to two old communications regarding a QCD treatment of nuclear SFs
in the single gluon exchange PWIA approximation. That approach leads
to a generalized convolution of distribution functions much like Eq.
(\ref{a2a}), with a simple correspondence between the featuring
quantities in the two representations \cite{jaffe,west}.

The above approach can actually be extended beyond single gluon
exchange, and one shows that at least some higher order QCD
corrections can still be accommodated in a convolution \cite{rtjw}.
In fact, formally the same expression (\ref{a2a}) holds in both a
hadronic and a QCD representation for $F^A$, provided one
re-interprets $f^{A,PN}$ in the latter as the distribution function
of (centers of) nucleons in the target. It is not likely that there
are significant QCD contributions which cannot be accommodated in a
convolution. At the end of Section II we recalled and actually
compared EMC ratios, calculated in the hadron and in a parton
representation.

The availability of planned Rosenbluth-separated data naturally
simplifies the analysis, but will probably not resolve the exposed
problems, as long as the scatter of neighboring points is much
larger than the accuracy of each  point. For instance $\alpha_n$,
extracted from Eq. (\ref{a13}) will remain sensitive to the input.

Use of the the same analyzing tools as before, indicates that the
new JLab  do not confirm previous conclusions, and--related to the
above, one cannot extract statistically significant information on
the neutron, in contrast to the apparent success, previously
obtained from OD. The resolution of this difficulty is clearly
highly desired.

\section {Acknowledgements.}

The authors are grateful to John Arrington, who provided us with all
measured total inclusive cross section, being final or not, and made
useful remarks. Thanks are equally due to Gianni Salme, who provided
the 3 sets of $A=3$ Spectral Functions, which enabled a realistic
calculation of SFs for the $A=3$ system.

\section {Appendix.}

We start with  a QCD prediction for the lowest moment
$M_0(Q^2)=\int_0^1 dx F_2^p(x,Q^2)=0.1471$ of a nuclear SF in the
Bjorken limit with $N_f=6$ contributing flavors. For any nuclear
target a similar moment can be computed, given a parton
representation of $F_2^A$. For several $A$ and finite
$Q^2=(2.5-10.0)$ GeV$^2$, we found values up to 5-6$\%$ lower than
for a $p$ \cite{rtpart}.

We now add the moments of the computed $F_2^{^{^3}He}$ to previously
reported results for $\langle N\rangle$, D, $^4$He, C, Fe in Table
I, \cite{rtpart}. For low $Q^2\approx 2.5$ GeV$^2$ the moment of
$^3$He is 16$\%$ higher than for a $p$. That moment rapidly
decreases with increasing $Q^2$, specifically to 0.1476 for
$Q^2=10\,$GeV$^2$, close to the Bjorken limit. The same for $^3$H is
substantially closer to the previously computed moments of other
light and medium-weight targets. As Figs 5. illustrate the above is
due to the differences of $f_{p,n}$ and of each distribution
function for $^3$He and $^3$H. The same causes the differences in
the predicted EMC ratios $\mu^{A=3}$.

We return to $f^A$ in Eqs. (2.2), (2.3) which we termed the SF of a
fictitious nucleus composed of point-nucleus or, alternatively, a
distribution function for the centers of nucleons in a nucleus. QCD
is clearly not applicable to those artifacts, no matter how high
$Q^2$ is. Their norm ${\cal N}$, very close to 1, differs from
$M_0^A$ of physical nuclei.

In Section IIB we mentioned that for either choice of $NN$
interaction B2 or B2+B3, the norm of the lowest order part ${\cal
N}_x(Q^2)=\int_0^3 dx f_{p,n}^{(0);A=3}(x,Q^2)$ equals 1 within a
few parts per mille: More precisely, for $Q^2=(2.5-10.0)$ GeV$^2$,
${\cal N}_x$ has a minute slope $\approx 0.0024/$GeV$^2$. The same
holds for ${\cal N}_y(q)=\int_{-q/2}^{q_{max}} dy \phi^{(0)}(q,y)$.

The above result for finite $Q^2$ is not self-understood. Only in
the $Q^2\to\infty$ limit can one easily derive ${\cal N}$=1.0. For
finite $Q^2$ the deviations of the norm from 1 and the minute slope
of those deviations as function of $Q^2$ are not due to numerical
inaccuracies. We note that a total disregard of the missing energy
and momentum yields $f(x,Q^2)=\delta(x-1)$.

It is interesting to observe that the relevant missing energies
appear restricted to $-|B_D|\lesssim E/M\lesssim 0.012$, which
dominates the underlying Spft $S_{p,n}^{A=3}(k,E)$ (negative values
occur only for a D-spectator). To a less extreme extent the same
holds for the missing momentum: $k/M\lesssim 0.2$. Re-instating
finite small missing energies and momenta produces distribution
functions with finite $Q^2$-dependent widths.

The above emphasizes $A=$3 nuclei, but several points hold for
general $A$. For the D a norm 1 is trivial, but a similar
observation can, and has been made for $^4$He, for which the
underlying Spft is fairly accurately known \cite{vkr}. For heavier
nuclei with models for $\rho_2^A$ (see Ref. I, Eqs. (9),(10)), the
norm ${\cal N}(f)$, when necessary, has been adjusted to 1.

Finally, the above holds for the lowest order part $f^{(0)}$. For
the reasons mentioned, it is not evident that, when  including FSI
contributions, one should apply a small re-normalization correction,
due to the relatively small $f^{A=3;(1)}<<f^{A=3;(0)}$. We have not
done so in our calculations.

\newpage
\begin{table}
{\small

\protect\caption {Values of $F_2^D(x,Q^2)$, extracted and computed
from ND, and the same from CL \cite{osip3} for a chosen $x$-range
(columns 1). Columns 2 give $Q^2(x,\theta)$ for that $x$  range and
$\theta=18^{\circ},22^{\circ}$, respectively $\theta=32^{\circ},
40^{\circ}$. Columns 3,4,5, are  for ND $F_2^{D,extr}, F_2^{D,calc},
F_2^{D, CL}$ for given $x$ and the above two pairs of angles .}

\begin{tabular} {|c|c||c|c|c||c||c|c|c|}
\hline

$x$ & $Q^2(18^{\circ})$ & $F_2^{D;R}(18^{\circ})$
&$F_2^{D;calc}(18^{\circ})$ & $F_2^{D:CL}(18^{\circ})$ &
$Q^2(26^{\circ})$ & $F_2^{D;R}(26^{\circ})$
&$F_2^{D,calc}(26^{\circ})$&
$F_2^{D;CL}(26^{\circ})$  \\
\hline
  0.5 &2.03 & 0.153  & 0.154  & 0.156  & ---- & -----  & -----  & -----  \\
  0.6 &2.17 & 0.0985 & 0.102  & 0.105  & 3.31 & 0.0884 & 0.0910 & 0.0094 \\
  0.7 &2.28 & 0.0555 & 0.0530 & 0.0559 & 3.56 & 0.0522 & 0.0542 & 0.0556 \\
  0.8 &2.36 & 0.0387 & 0.0333 & 0.0464 & 3.79 & 0.0241 & 0.0238 & 0.0272 \\
  0.9 &2.43 & 0.0225 & 0.0200 & 0.0171 & 3.98 & 0.0109 & 0.0108 & 0.0126 \\
  1.0 &2.51 & 0.0344 & 0.0369 & 0.0406 & 4.15 & 0.0110 & 0.0123 & 0.0129 \\
  1.1 &2.56 & 0.0086 & 0.0112 & 0.0097 & 4.29 & 0.0022 & 0.0026 & 0.0022 \\

\hline

$x$ &  $Q^2(32^{\circ})$ & $F_2^{D;R}(32^{\circ})$ &
$F_2^{D;calc}(32^{\circ})$ & $F_2^{D;CL}(32^{\circ})$ &
$Q^2(40^{\circ})$ & $F_2^{D;R}(40^{\circ})$ &
$F_2^{D;calc}(40^{\circ})$ &
$F_2^{D;CL}(40^{\circ})$\\
\hline
  0.4 & --- &  ---    &  ---  &   ---  & 3.37 & 0.184  & 0.174  & 0.187  \\
  0.5 & --- &  ---    &  ---  &   ---  & 4.02 & 0.125  & 0.124  & 0.128  \\
  0.6 &3.95 & 0.0801 & 0.0810 & 0.0775 & 4.59 & 0.0740 & 0.0750 & 0.0785 \\
  0.7 &4.33 & 0.0471 & 0.0480 & 0.0450 & 5.09 & 0.0404 & 0.0415 & 0.0442 \\
  0.8 &4.66 & 0.0204 & 0.0210 & 0.0225 & 5.58 & 0.0186 & 0.0182 & 0.0212 \\
  0.9 &4.95 & 0.0089 & 0.0083 & 0.0096 & 5.99 & 0.0065 & 0.0058 & 0.0088 \\
  1.0 &5.23 & 0.0055 & 0.0064 & 0.0053 & 6.39 & 0.0034 & 0.0035 & 0.0035 \\
  1.1 &5.49 & 0.0008 & 0.0010 & 0.0059 & 6.72 & 0.0005 & 0.0006 & ---    \\

\hline

\end{tabular} }
\end{table}

\begin{table}
{\small


\caption{ $\alpha_n$ extracted from inclusive scattering  on D
\cite{osip3}. Columns give group of values of $Q^2$, around
$Q^2=2.50, 3.34, 4.15, 5.24$ GeV$^2$, which correspond to the values
$Q^2(\theta=18^{\circ }, 22^{\circ}, 26^{\circ}, 32^{\circ}; x=1)$
Columns 2-3 are the $x-$range of  points around QEP and number of
selected points $n$. The last column gives the average over the
given $x$-range of $\langle\alpha_n(Q)\rangle$, the reduced magnetic
FF of the $n$ and the error of the mean.}

\begin{tabular} {|c|c|c|c|}
\hline
$Q^2$ [GeV]$^2$ & $x$-interval & $n$ &
$\langle\alpha_n\rangle \pm
\delta \alpha_n$
\\

\hline

2.425 & 0.9235-1.0405 & 12 & 1.0005 $\pm$ 0.0334 \\
2.475 & 0.9235-1.0405 & 12 & 0.9837 $\pm$ 0.0310 \\
2.525 & 0.9235-1.0315 & 12 & 1.0020 $\pm$ 0.0294 \\
2.575 & 0.9235-1.0315 & 10 & 1.0488 $\pm$ 0.0324 \\
\hline
3.275 & 0.9415-1.0225 &  9 & 0.9752 $\pm$ 0.0344 \\
3.325 & 0.9595-1.0225 &  6 & 0.9917 $\pm$ 0.0475 \\
3.375 & 0.9325-1.0225 &  7 & 0.9720 $\pm$ 0.0448 \\
\hline
4.075 & 0.9685-1.0675 &  9 & 0.9822 $\pm$ 0.0430 \\
4.125 & 0.9775-1.0855 &  9 & 0.9614 $\pm$ 0.0497 \\
4.175 & 0.9685-1.0855 & 11 & 0.9415 $\pm$ 0.0354 \\
4.225 & 0.9775-1.0765 &  8 & 0.9804 $\pm$ 0.0309 \\
\hline
5.075 & 0.9865-1.0675 &  6 & 0.9237 $\pm$ 0.0593 \\
5.175 & 0.9685-1.0585 &  6 & 0.9001 $\pm$ 0.0350 \\
5.275 & 0.9865-1.0765 &  5 & 0.8753 $\pm$ 0.0625 \\
5.375 & 0.9685-1.1035 &  7 & 0.9145 $\pm$ 0.0378 \\

\hline

\end{tabular}
\label{Table II} }
\end{table}

\renewcommand\thefigure{1a}

\begin{figure}[p]
\includegraphics[scale=.8]{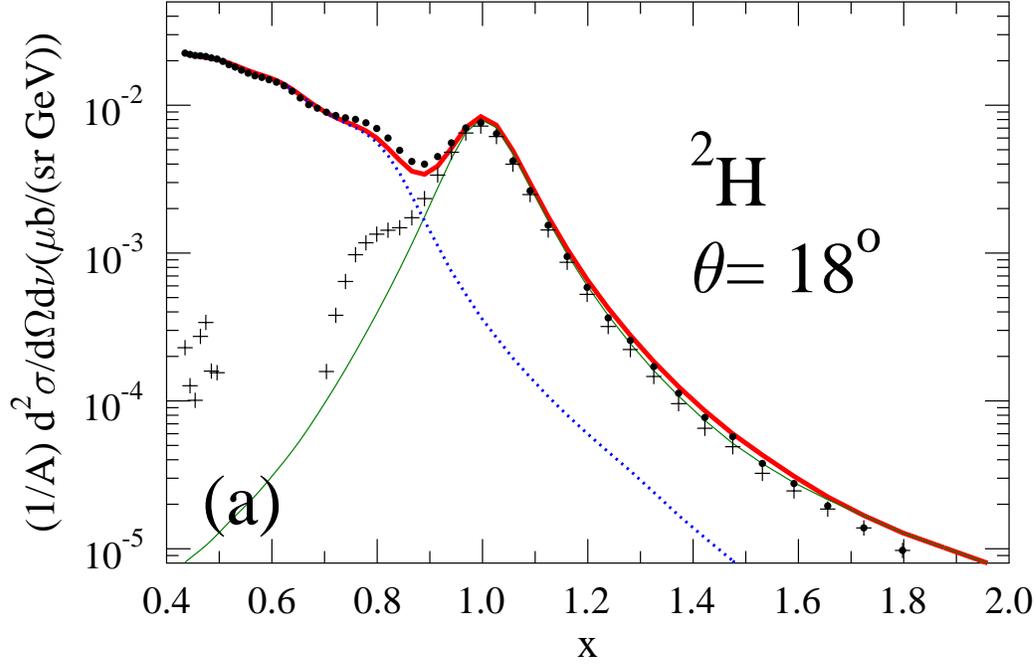}
\vspace{-12cm} \caption{$\sigma^{tot}$ for inclusive scattering of
$E=5.76$ GeV electrons on D; $\theta=18^{\circ}$. Heavy dots are
data without error bars. Small dots (blue) and thin line (green) are
NI$^{calc}$ and NE$^{FF}$, Eqs. (\ref{a2b}), (\ref{a3a}),
(\ref{a3b}). Heavy line (red) is their sum. Crosses are NE$^{extr}$=
data -- NI$^{calc}$, Eq. (\ref{a4}). Missing crosses indicate data
$\le \rm {NI}^{calc}$. }
\end{figure}

\renewcommand\thefigure{1b}

\begin{figure}[p]
\includegraphics[scale=.8]{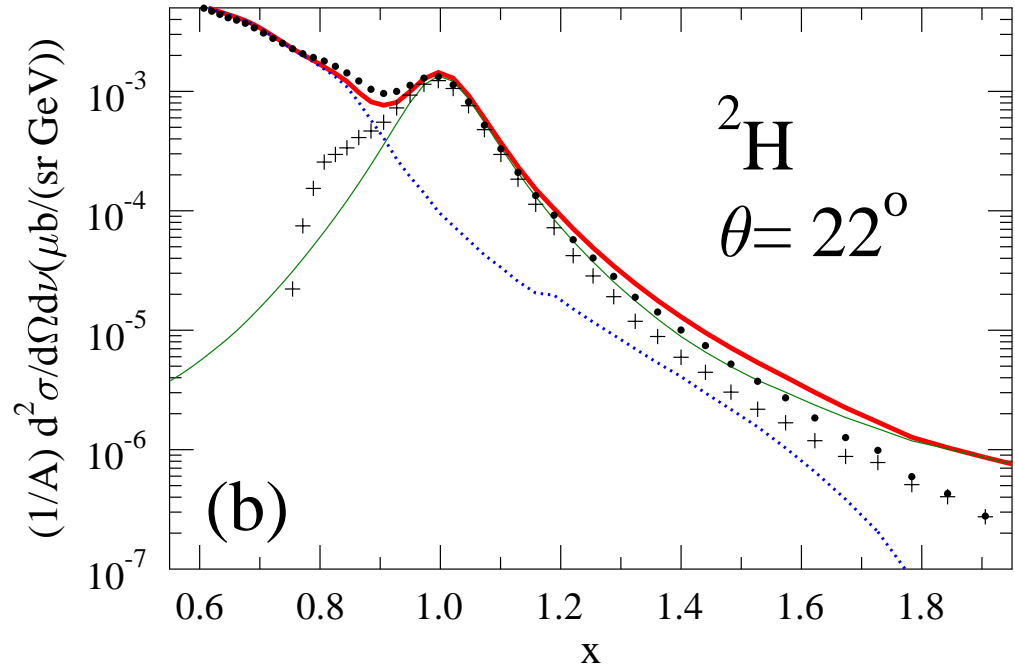}
\vspace{-12cm}\caption{Fig. 1a for $\theta=22^{\circ}$.}
\end{figure}

\renewcommand\thefigure{1c}

\begin{figure}[p]
\includegraphics[scale=.8]{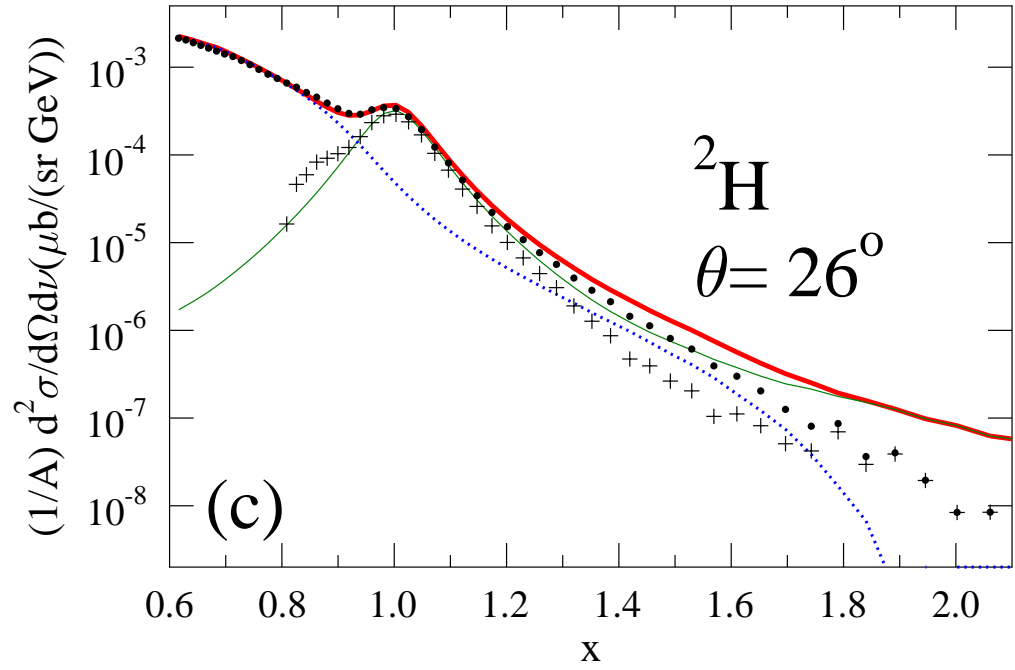}
\vspace{-12cm}\caption{Fig. 1a for $\theta=26^{\circ}$.}
\end{figure}

 \renewcommand\thefigure{1d}

\begin{figure}[p]
\includegraphics[scale=.8]{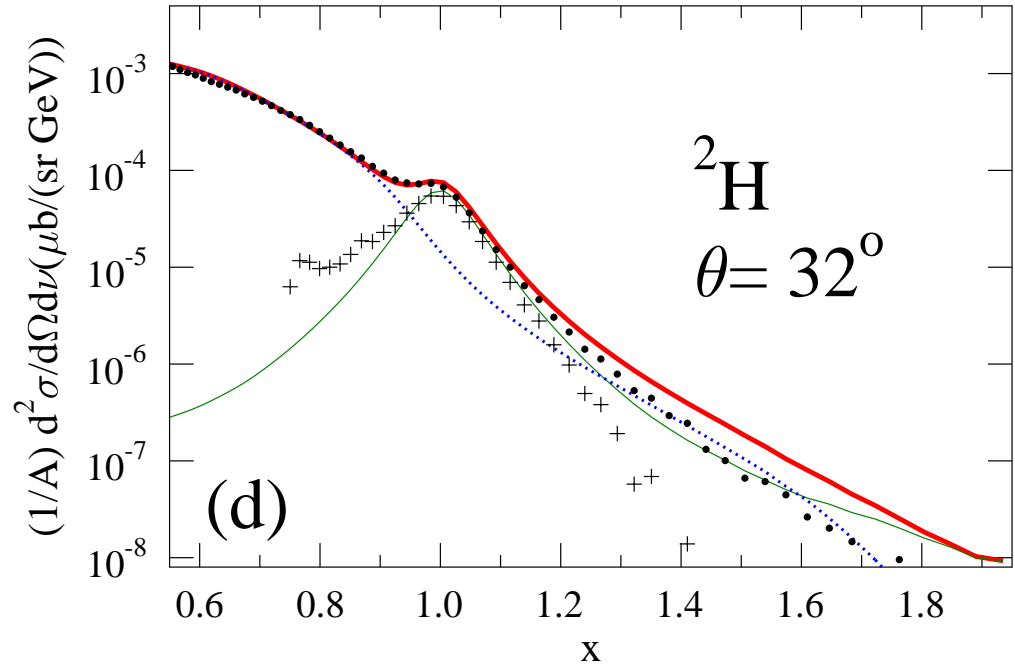}
\vspace{-12cm}\caption{Fig. 1a for $\theta=32^{\circ}$.}
\end{figure}

\renewcommand\thefigure{2a}

\begin{figure}[p]
\includegraphics[scale=.8]{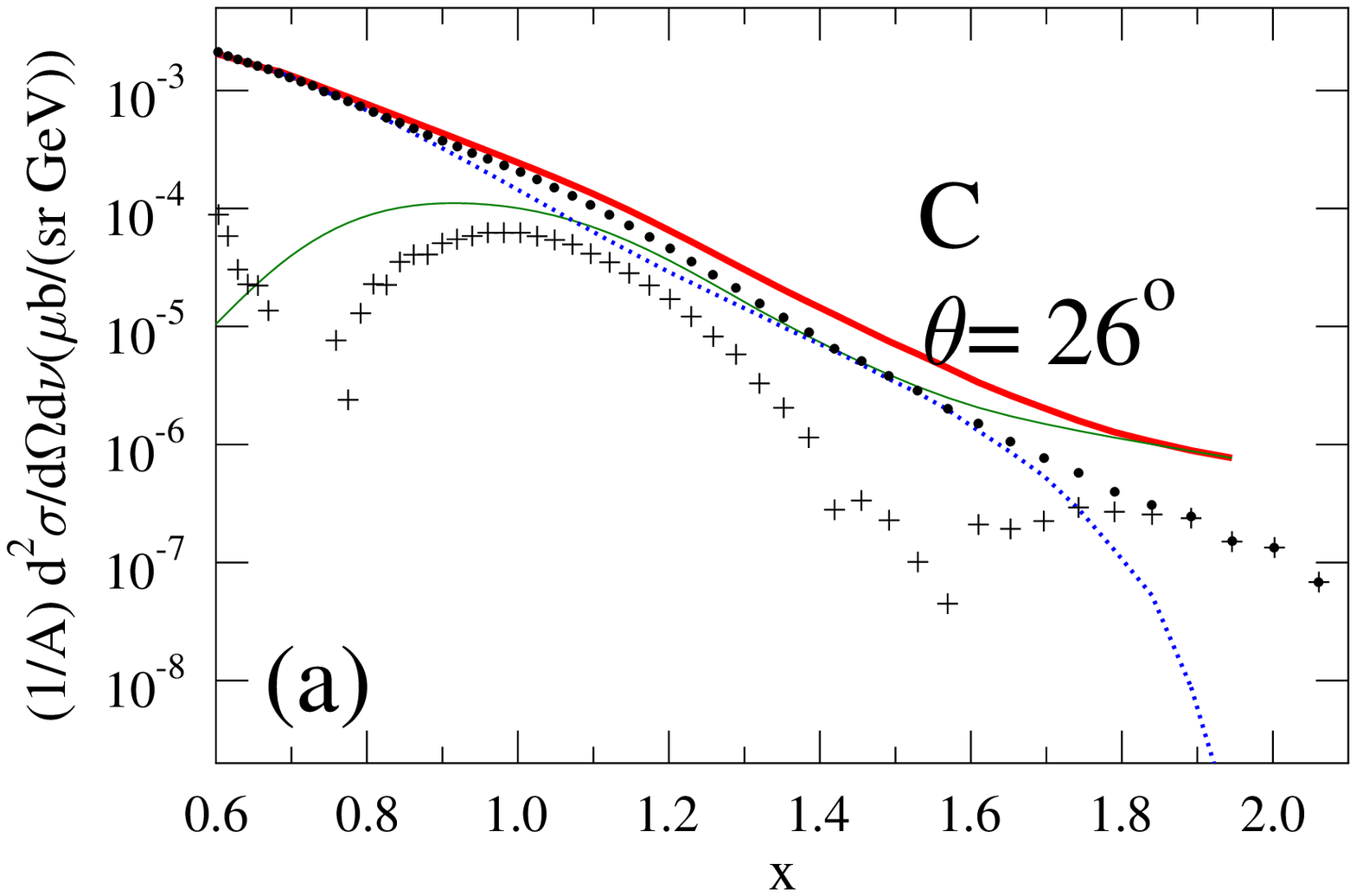}
\caption{Fig. 1c for C . }
\end{figure}

\renewcommand\thefigure{2b}

\begin{figure}[p]
\includegraphics[scale=.8]{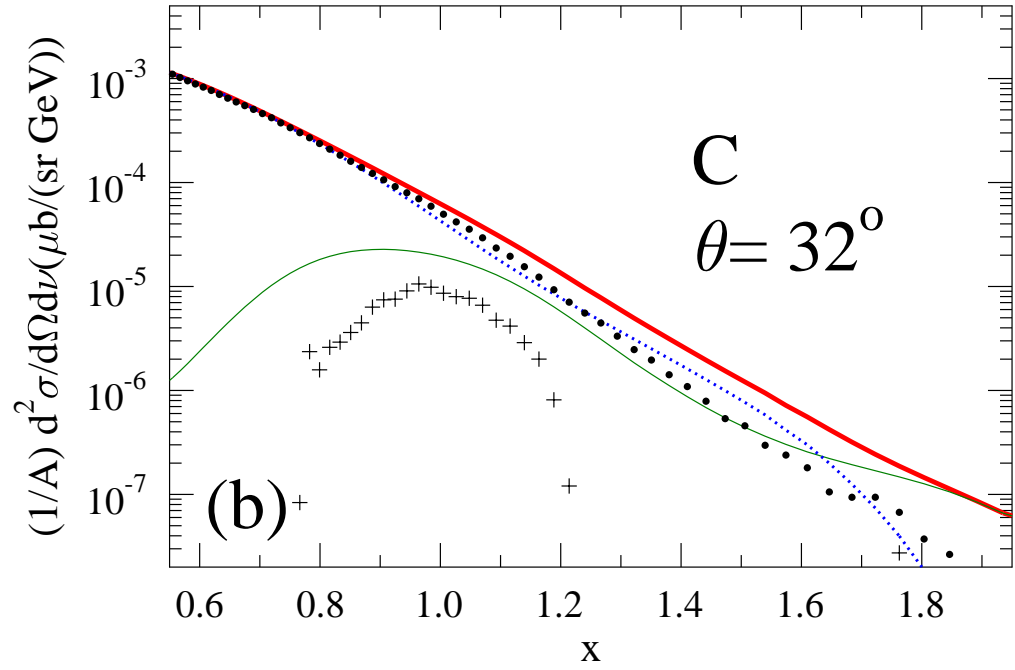}
\vspace{-12cm}\caption{Fig. 2a for $\theta=32^{\circ}$.}
\end{figure}

\renewcommand\thefigure{2c}

\begin{figure}[p]
\includegraphics[scale=.8]{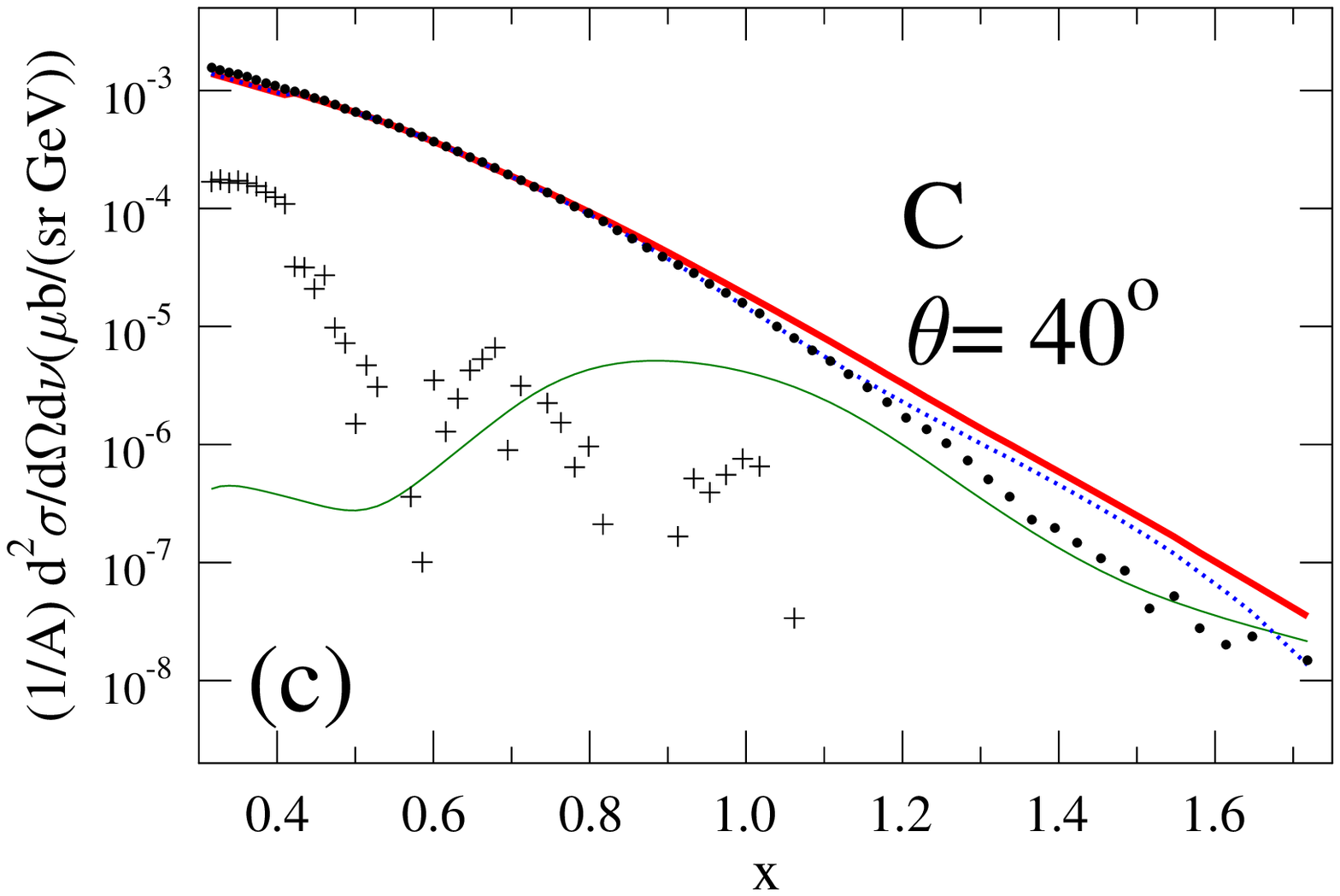}
\vspace{-12cm}\caption{Fig. 2a for $\theta=40^{\circ}$.}
\end{figure}

\renewcommand\thefigure{2d}

\begin{figure}[p]
\includegraphics[scale=.8]{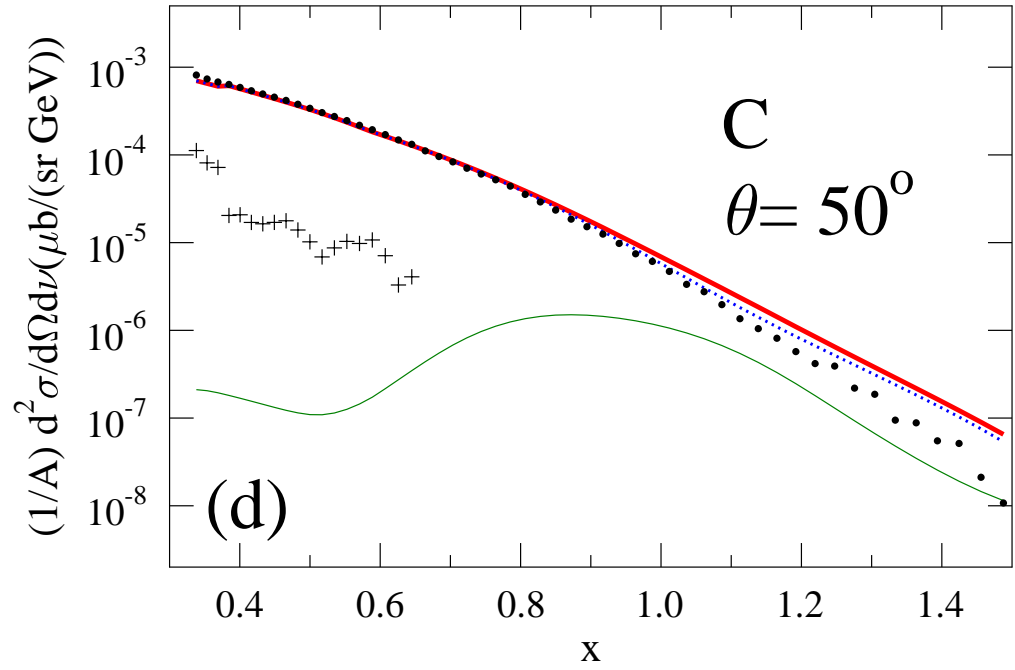}
\vspace{-12cm}\caption{Fig. 1a for $\theta=50^{\circ}$.}
\end{figure}

\renewcommand\thefigure{3a}

\begin{figure}[p]
\includegraphics[scale=.8]{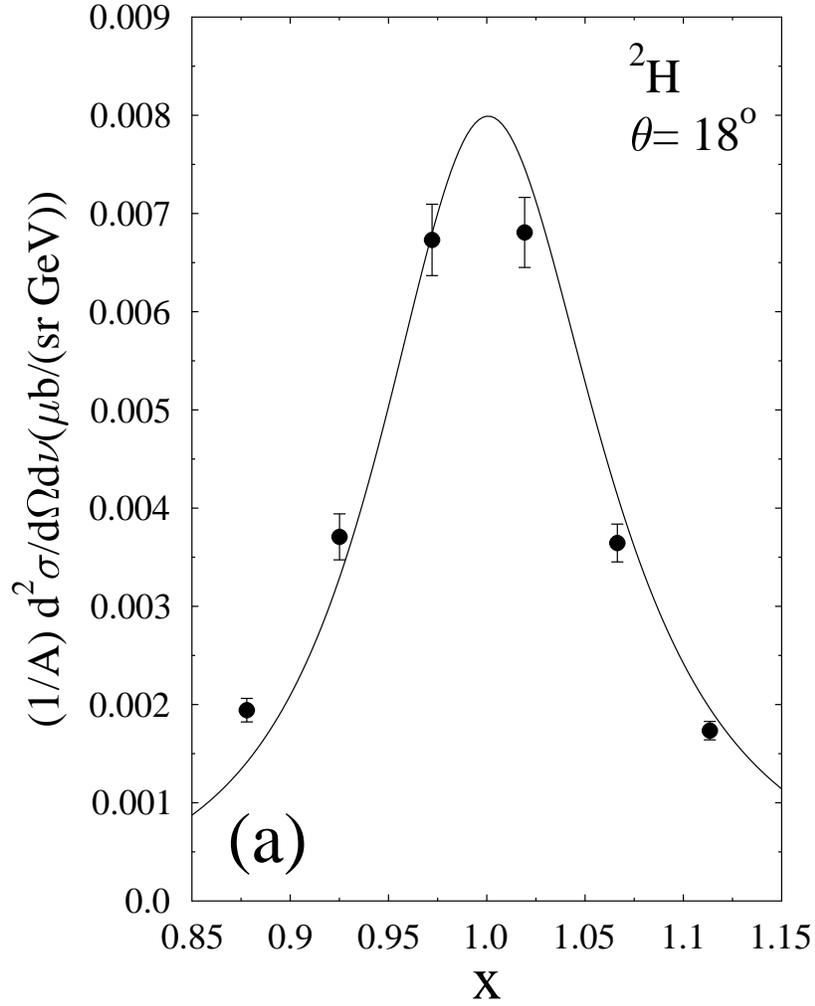}
\caption{ Linear plot of QE part of total inclusive cross section of
$E=5.76$ GeV  electrons on D, for $\theta=18^{\circ}$. Drawn line is
NE$^{FF}$, Eqs. (1.3),( 1.4). Filled circles are 'data' with total
error bars. }
\end{figure}

\renewcommand\thefigure{3b}

\begin{figure}[p]
\includegraphics[scale=.8]{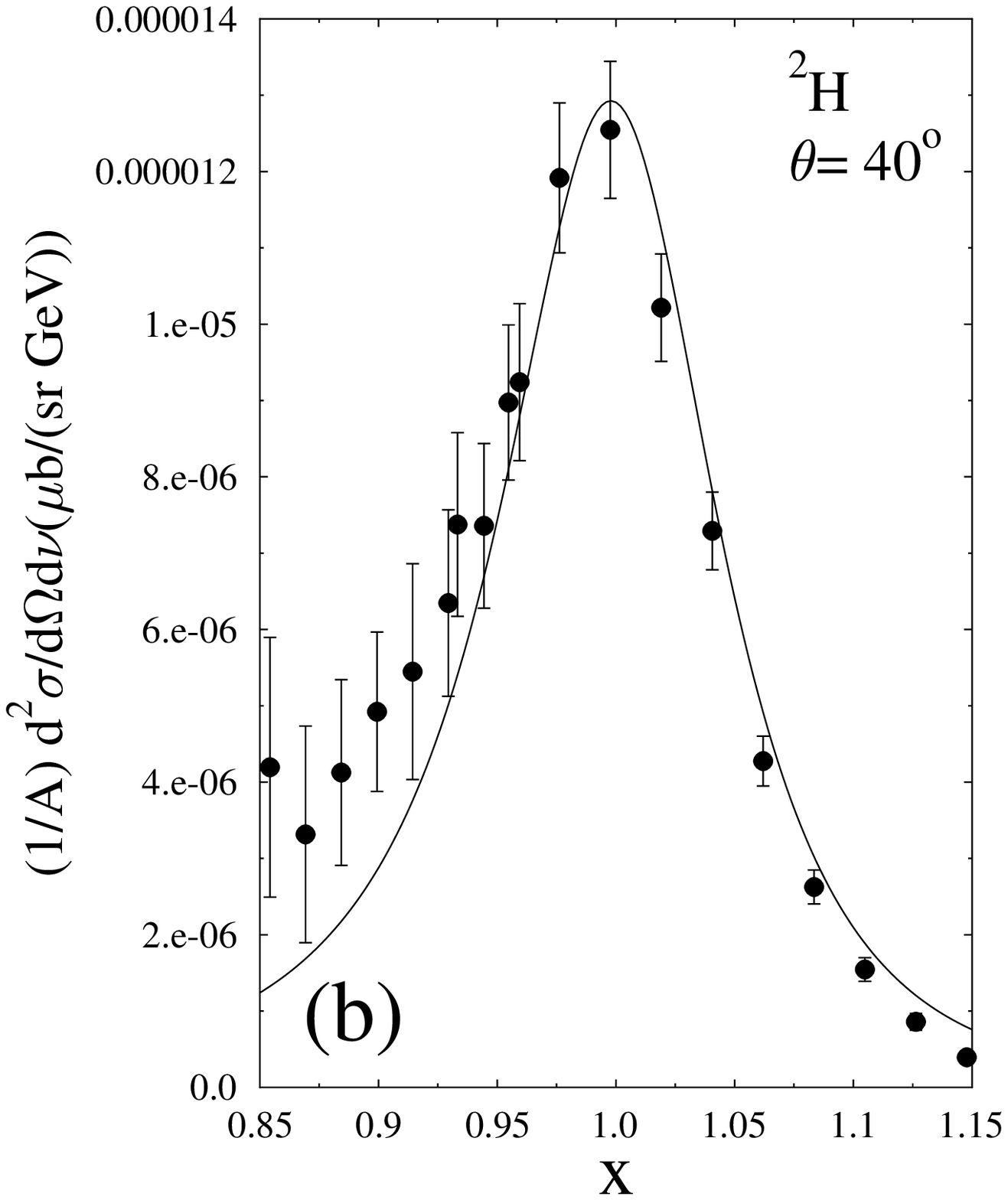}
\vspace{-2cm}\caption{Fig. 3a for $\theta=40^{\circ}$. }
\end{figure}

\renewcommand\thefigure{3c}

\begin{figure}[p]
\includegraphics[scale=.8]{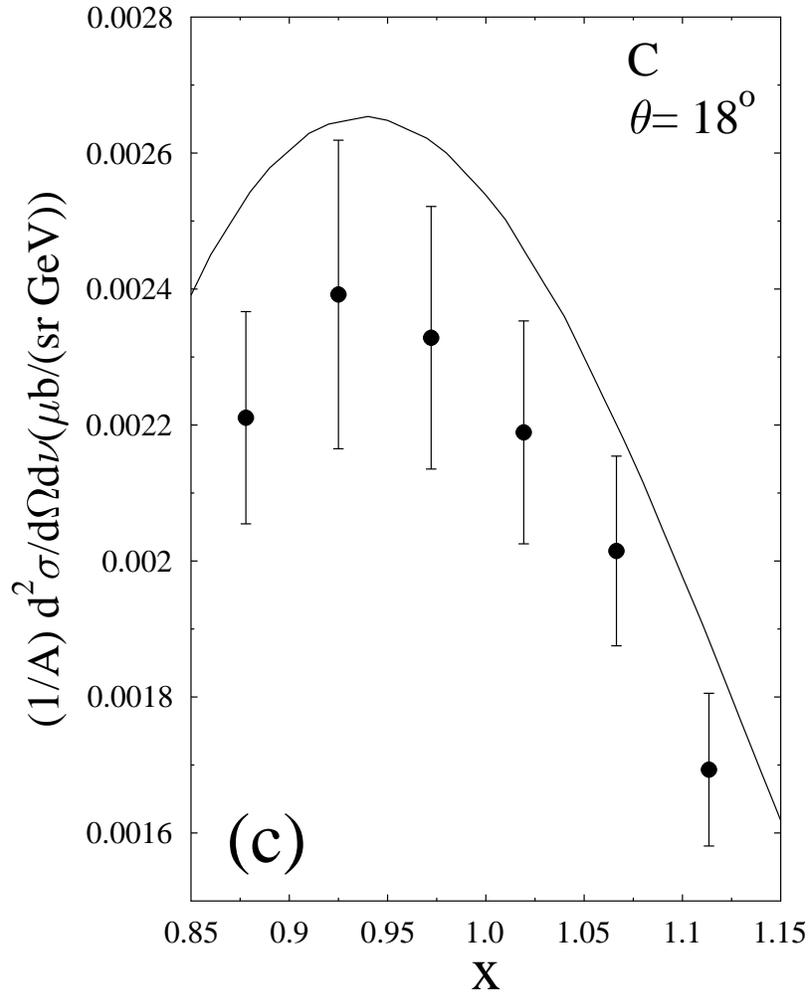}
\caption{Fig. 3a for C, $\theta=18^{\circ}$. }
\end{figure}

\renewcommand\thefigure{3d}

\begin{figure}[p]
\includegraphics[scale=.8]{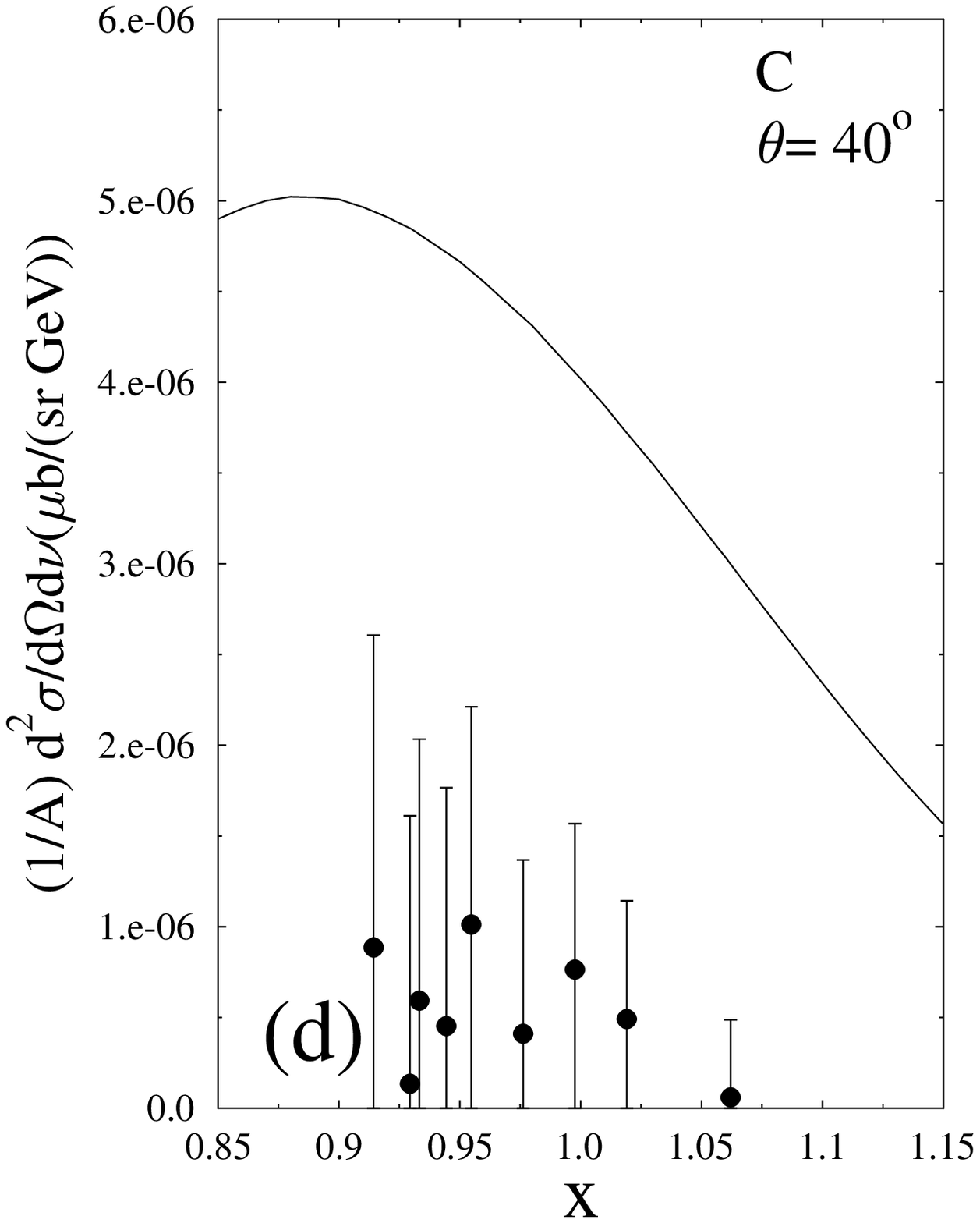}
\caption{Fig. 3c for $\theta=40^{\circ}$. }
\end{figure}

\renewcommand\thefigure{4a}

\begin{figure}[p]
\includegraphics[scale=.7]{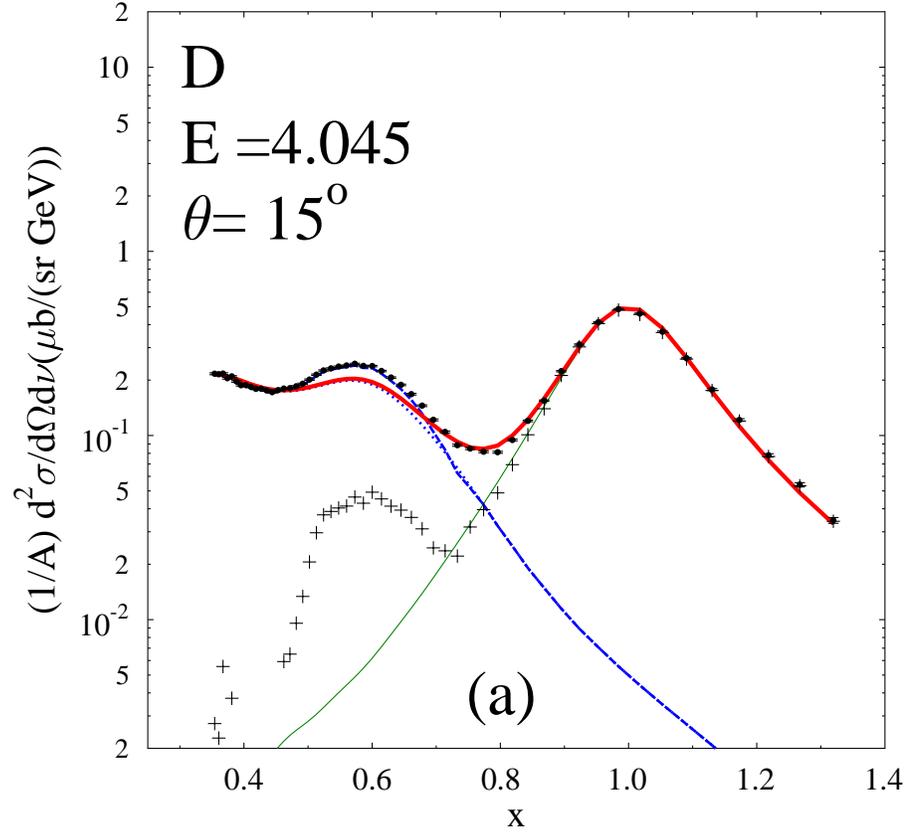}
\vspace{-2cm}\caption{ As Fig. 1a for $E$=4.045 GeV,
$\theta=15^{\circ}$ with same meaning of symbols and curves. Dashed
curve (blue) is for NI$^{emp}$, for which NE$^{FF}\approx$
NE$^{extr}$. (Fig. on log scale from Ref.\cite{rtv}). }
\end{figure}

\renewcommand\thefigure{4b}

\begin{figure}[p]
\includegraphics[scale=.8]{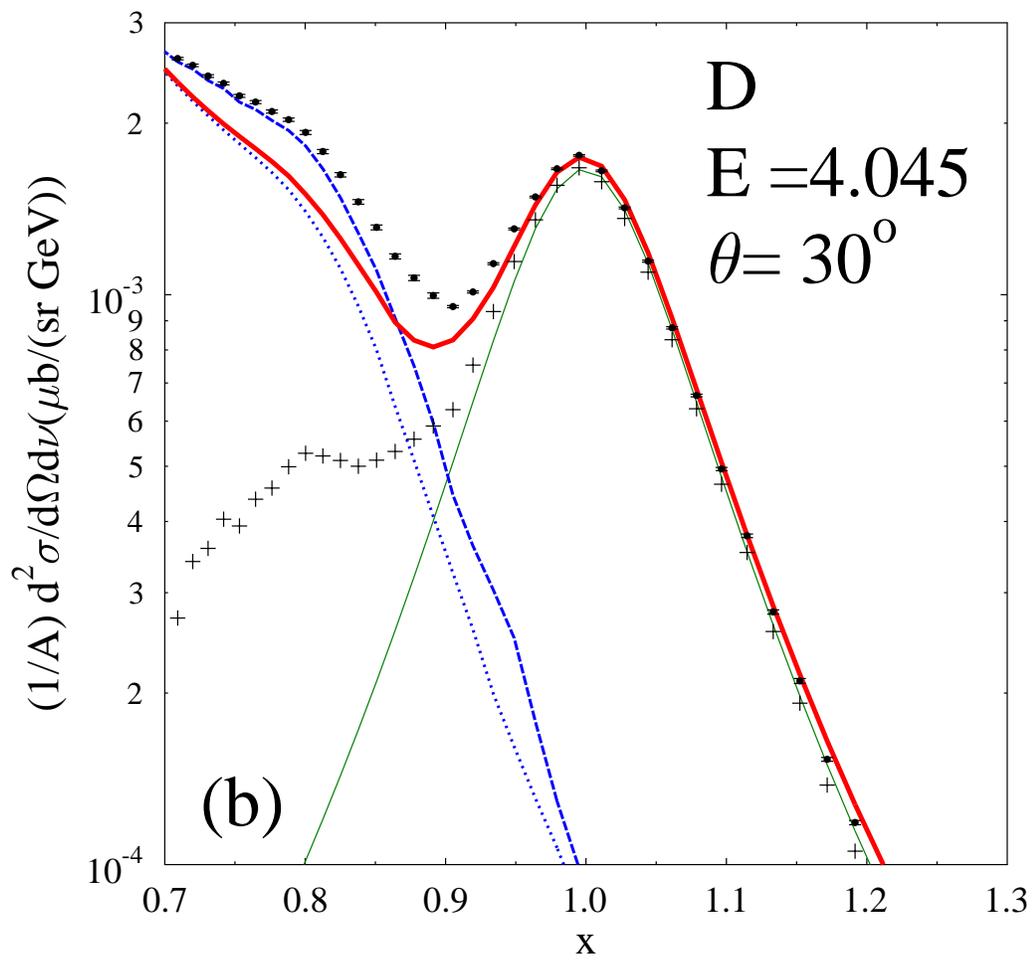}
\vspace{-2cm}\caption{Fig. 4a for $\theta=30^{\circ}$ . }
\end{figure}

\renewcommand\thefigure{4c}

\begin{figure}[p]
\includegraphics[scale=.8]{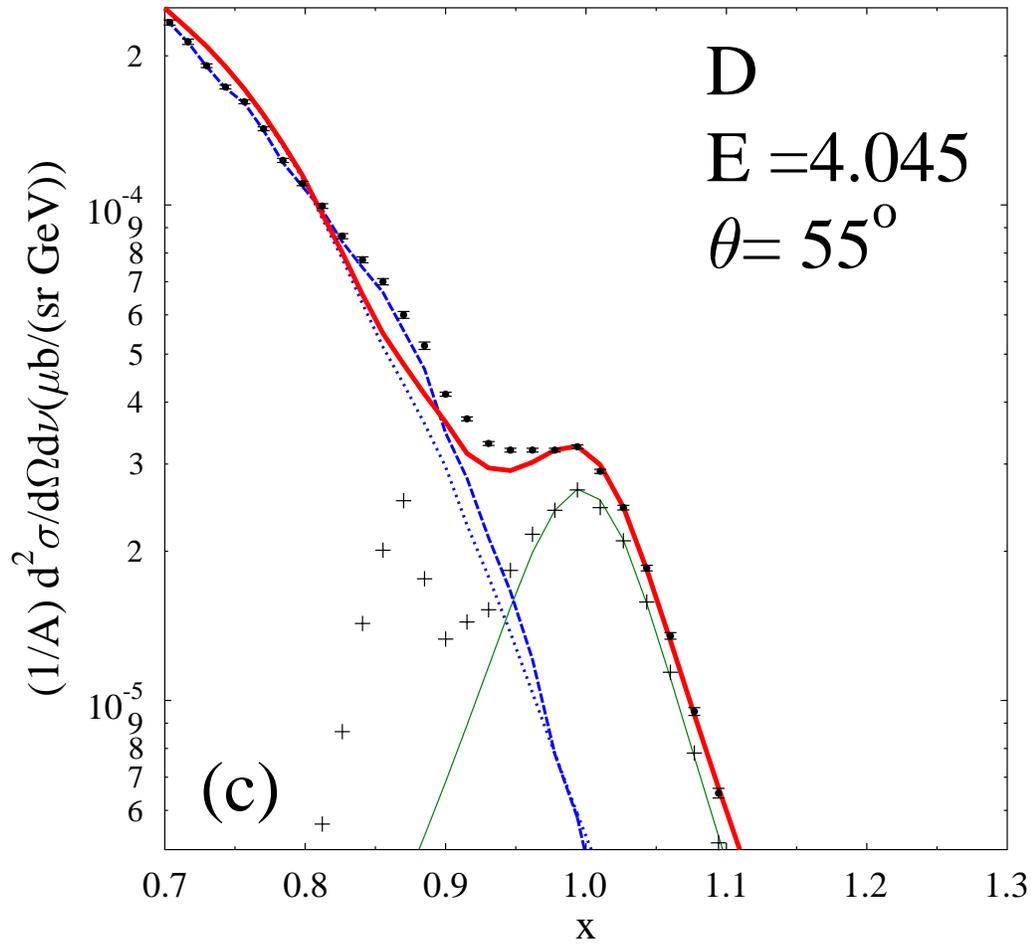}
\vspace{-2cm}\caption{Fig. 1a for $\theta=55^{\circ}$ . }
\end{figure}

\renewcommand\thefigure{5a}

\begin{figure}[p]
\includegraphics[scale=.8]{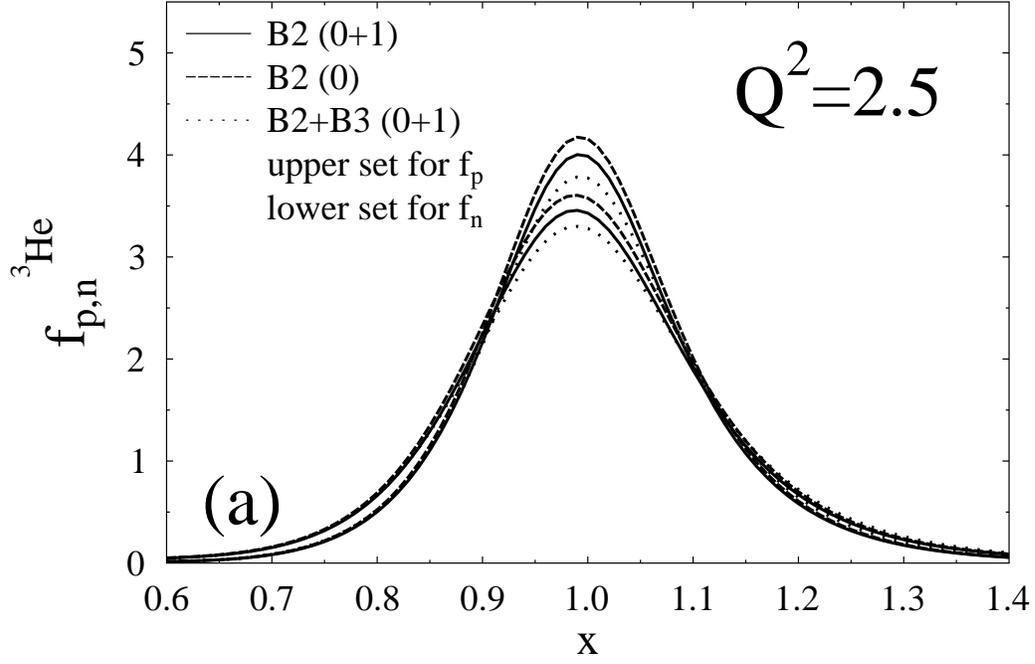}
\vspace{-12cm}\caption{ Distribution functions
$f_{p,n}^{{^3}He}(x,Q^2=5$ GeV$^2$). Drawn lines, dashes and dots
are for B2(0+1), B2(0) and B2+B3(0+1) . }
\end{figure}

\clearpage
\renewcommand\thefigure{5b}

\begin{figure}[p]
\includegraphics[scale=.8]{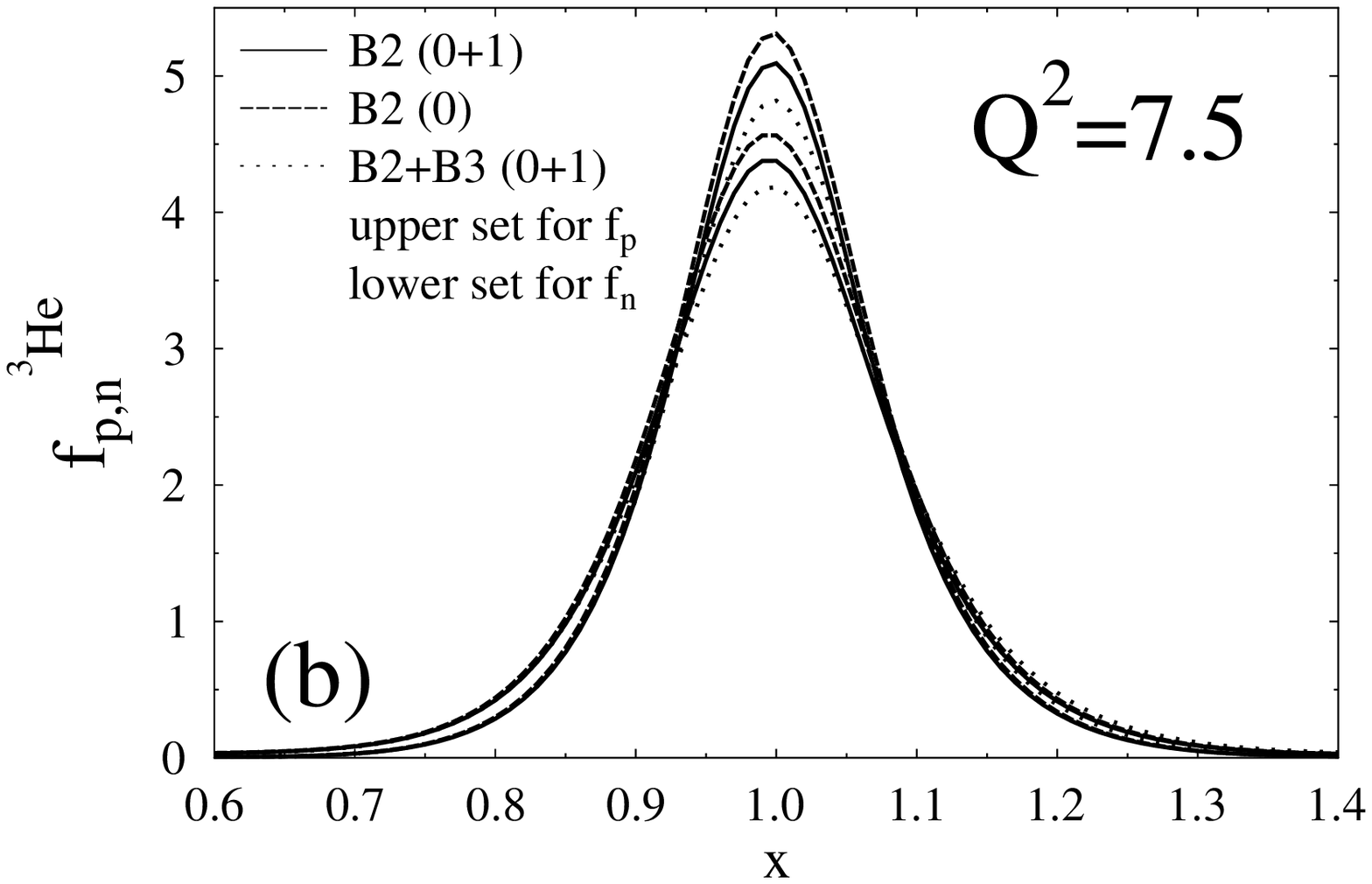}
\vspace{-12cm}\caption{Fig. 5a for $Q^2=7.5\,$GeV$^2$ .}
\end{figure}

\renewcommand\thefigure{5c}

\begin{figure}[p]
\includegraphics[scale=.8]{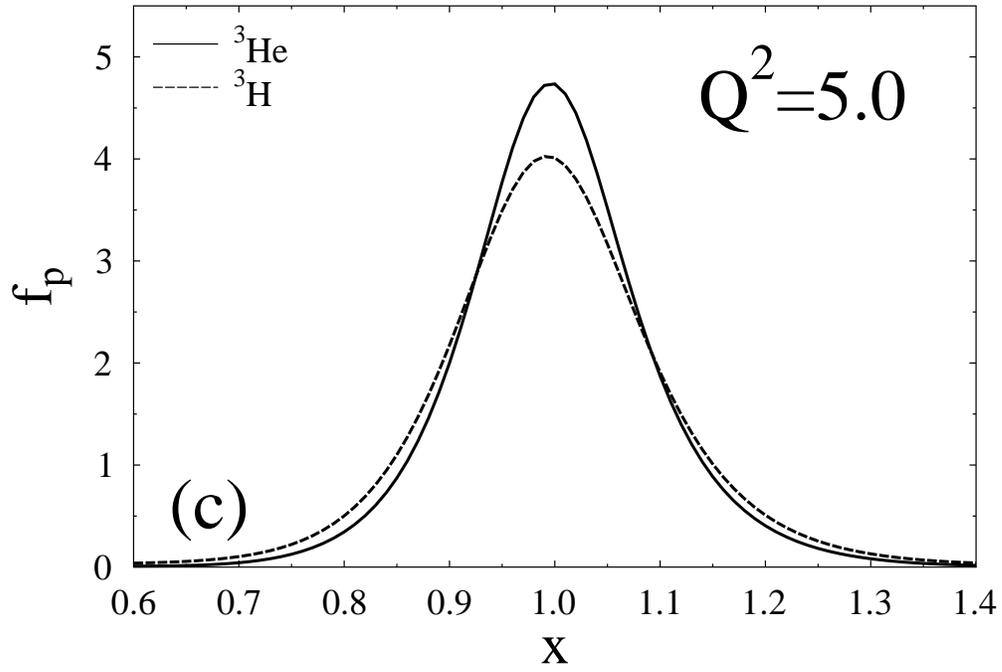}
\vspace{-12cm}\caption{Comparison of  $^3$He, $^3$H distribution
functions $f_p(x,Q^2=7.5)$, using B2(0+1). }
\end{figure}

\renewcommand\thefigure{6}

\begin{figure}[p]
\includegraphics[scale=.8]{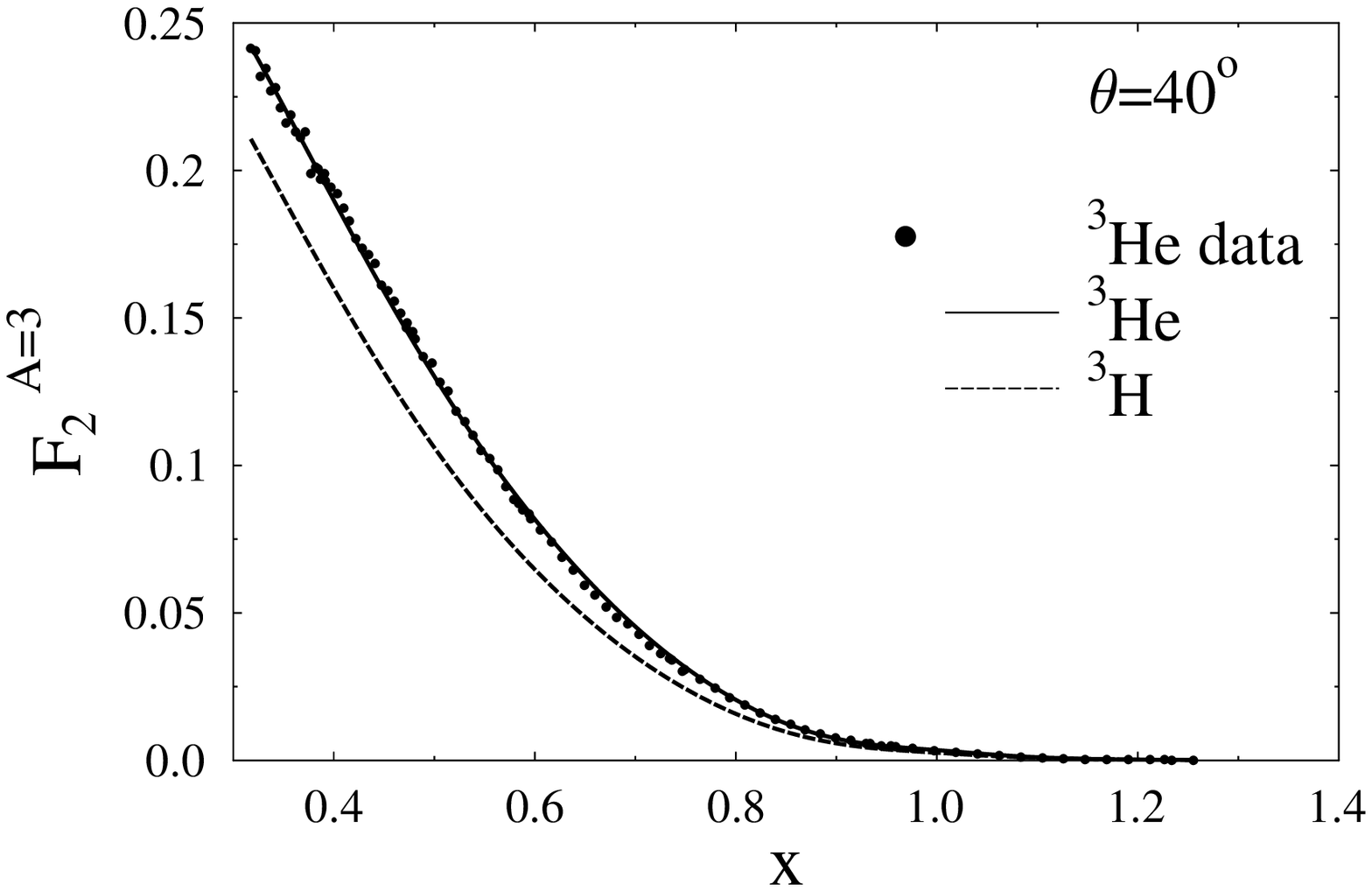}
\vspace{-12cm}\caption{Computed and extracted
$F_2^{^{3}He}(\theta=40^{\circ})$ for B2+B3 interactions. Lower
curve is for $F_2^{^{3}H}(\theta=40^{\circ})$ . }
\end{figure}
\clearpage

\renewcommand\thefigure{7a}

\begin{figure}[p]
\includegraphics[scale=.8]{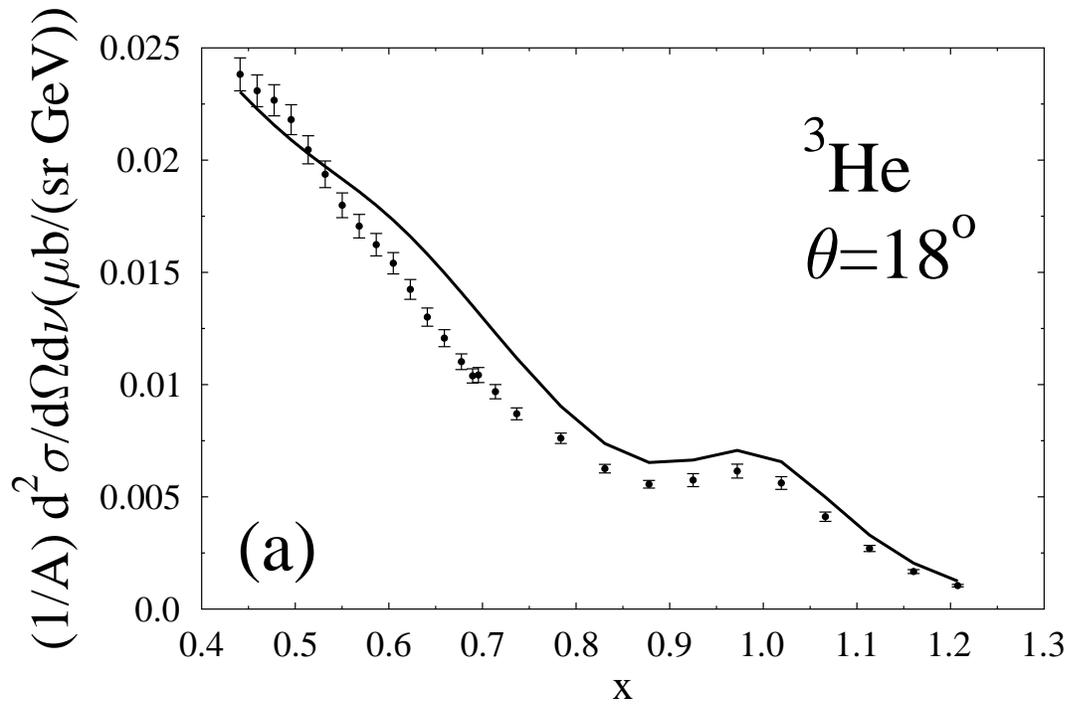}
\vspace{-12cm}\caption{Computed total inclusive cross section on
$^3$He for $\theta=18^{\circ}$. 'Data'  with error bars  are from
\cite{ag} . }
\end{figure}

\renewcommand\thefigure{7b}

\begin{figure}[p]
\includegraphics[scale=.8]{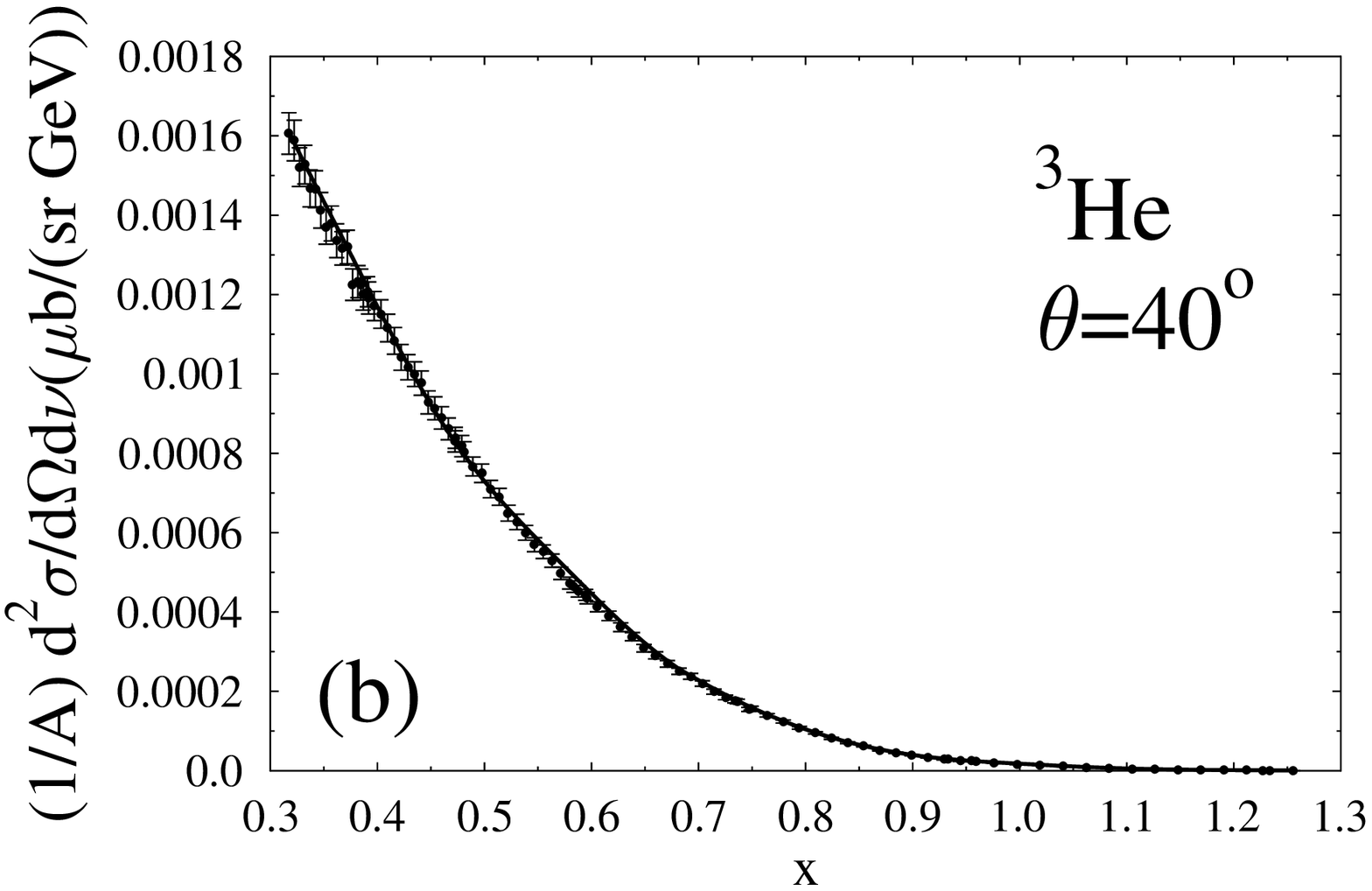}
\vspace{-12cm}\caption{Fig. 7a for $\theta=40^{\circ}$  . }
\end{figure}

\renewcommand\thefigure{8a}

\begin{figure}[p]
\includegraphics[scale=.8]{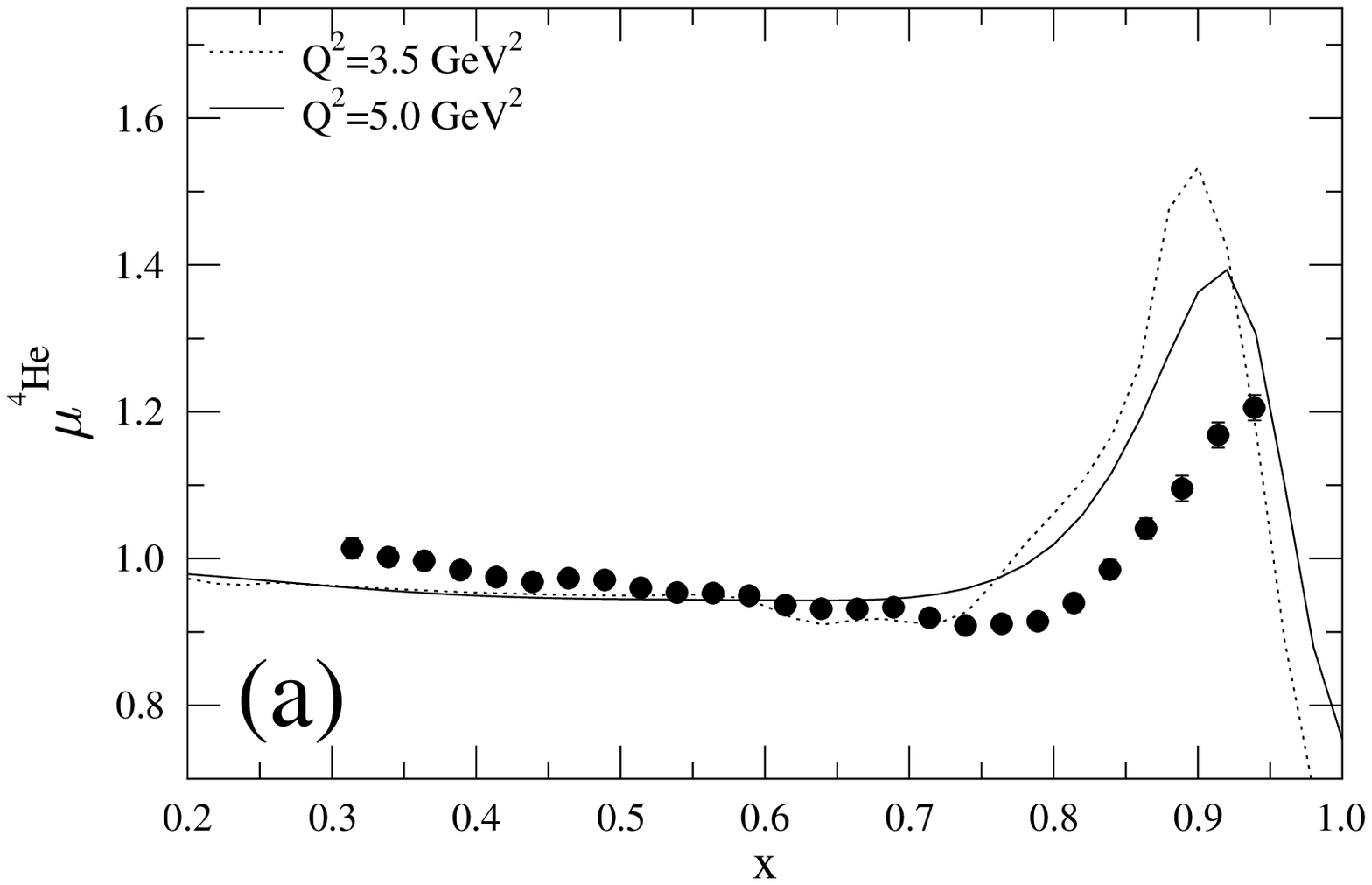}
\vspace{-12cm}\caption{ EMC data for $\mu^{^{4}He}(x \lesssim
1.0,\theta=40^{\circ}$) \cite{seely} and computed results for fixed
$Q^2=3.5, 5.0$ GeV$^2$ \cite{rtvemc}.}
\end{figure}

\renewcommand\thefigure{8b}

\begin{figure}[p]
\includegraphics[scale=.8]{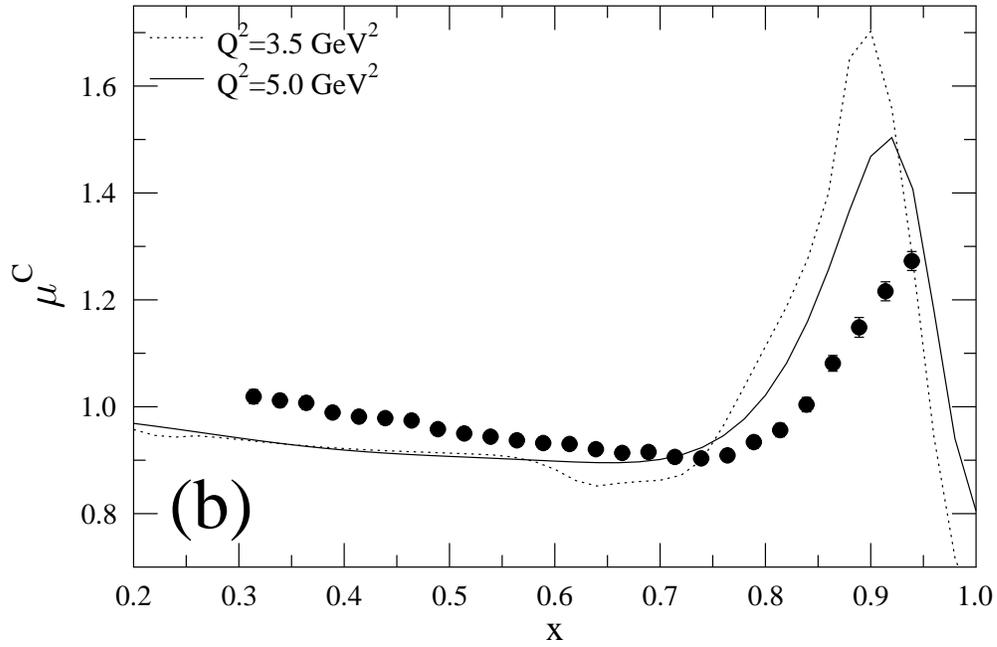}
\vspace{-12cm}\caption{ Fig. a for $\mu^C(x \lesssim
1.0,\theta=40^{\circ}$) \cite{seely}. }
\end{figure}

\renewcommand\thefigure{9a}

\begin{figure}[p]
\includegraphics[scale=.8]{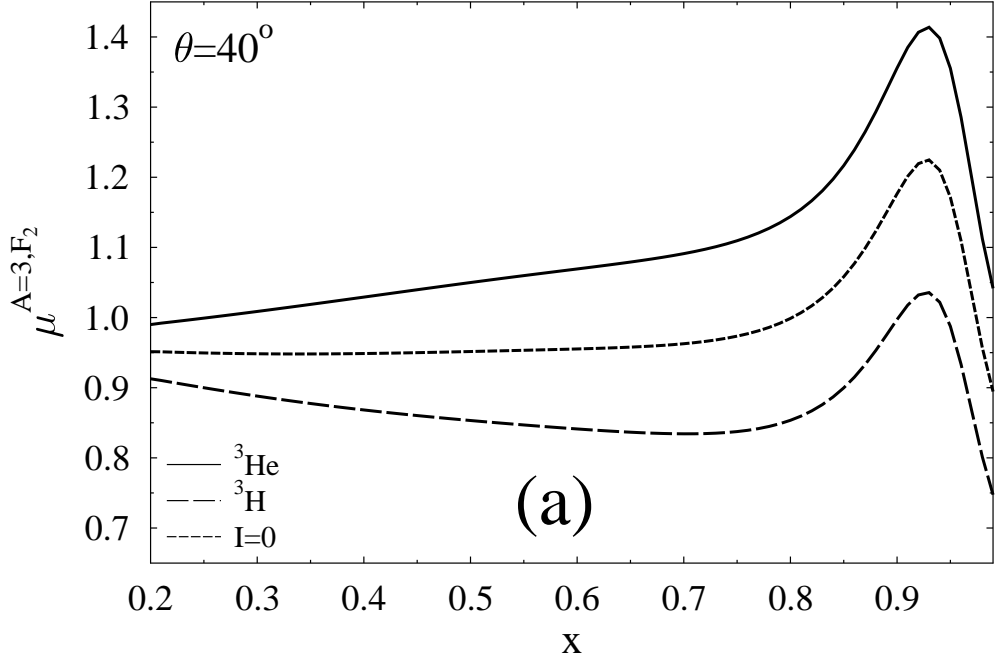}
\vspace{-12cm}\caption{ $\mu^{A=3;F_2}(\theta=40^{\circ})$ from
ratios of SFs $F_2$ for $^3$He and $^3$H (drawn line and long
dashes), and for the iso-scalar component (short dashes) . }
\end{figure}

\renewcommand\thefigure{9b}

\begin{figure}[p]
\includegraphics[scale=.9]{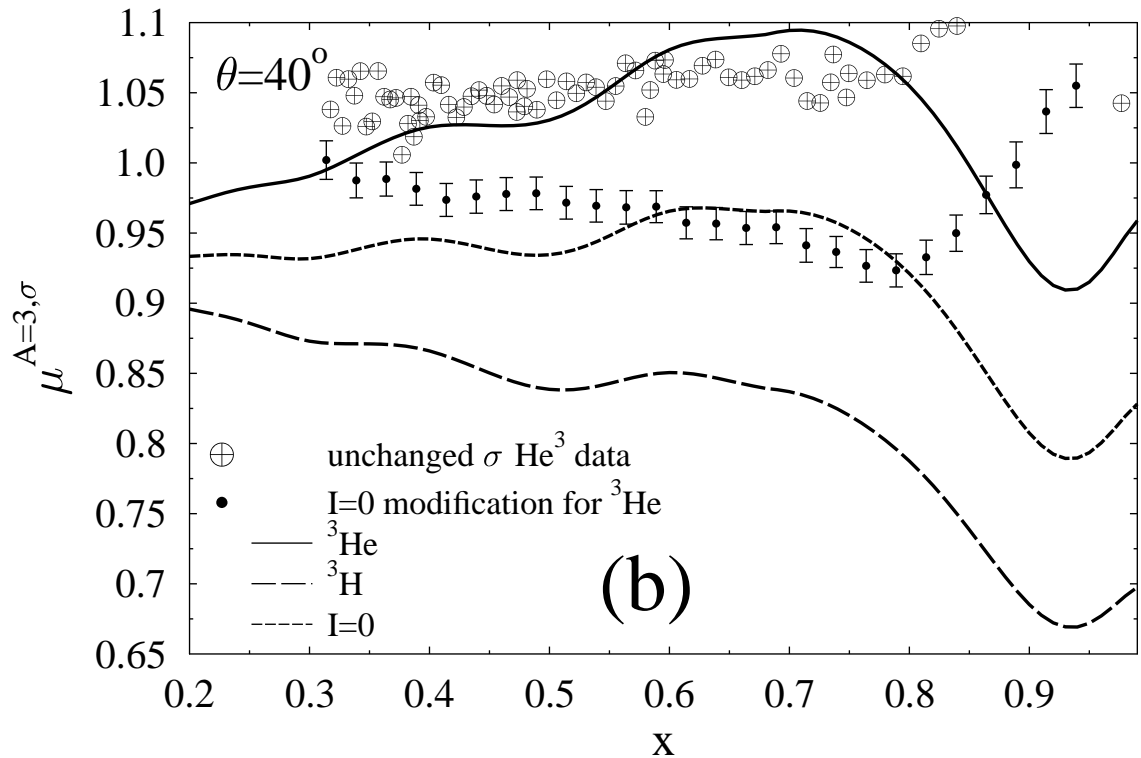}
\vspace{-12cm}\caption{ $\mu^{A=3;\sigma}(\theta=40^{\circ})$ from
ratios of cross sections. Curves as in Fig. 9a. Empty and full
circles are direct data for $^3$He, respectively manipulated ones
for a fictitious iso-scalar A=3 nucleus \cite{seely} .}
\end{figure}

\renewcommand\thefigure{10a}

\begin{figure}[p]
\includegraphics[scale=.8]{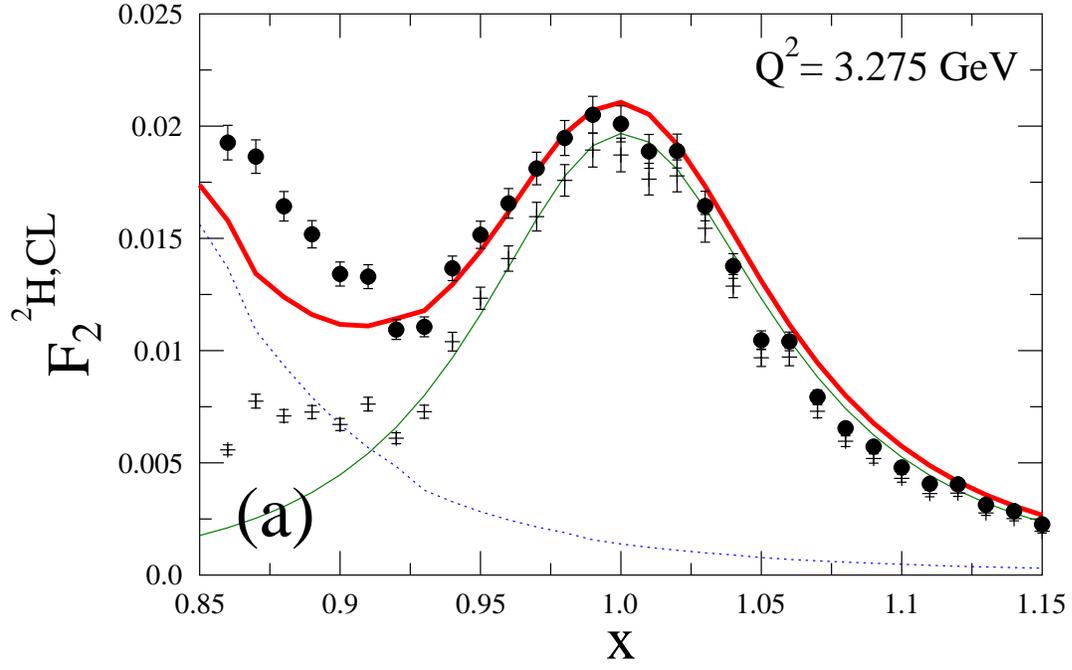}
\vspace{-12cm} \caption{ Filled circles  are data with error bars
for SF $F_2^D(x,Q^2=3.275\, $GeV$^2)$ from CL data\cite{osipd}.
Symbols and curves as in Fig. 1a . }
\end{figure}

\renewcommand\thefigure{10b}

\begin{figure}[p]
\includegraphics[scale=.8]{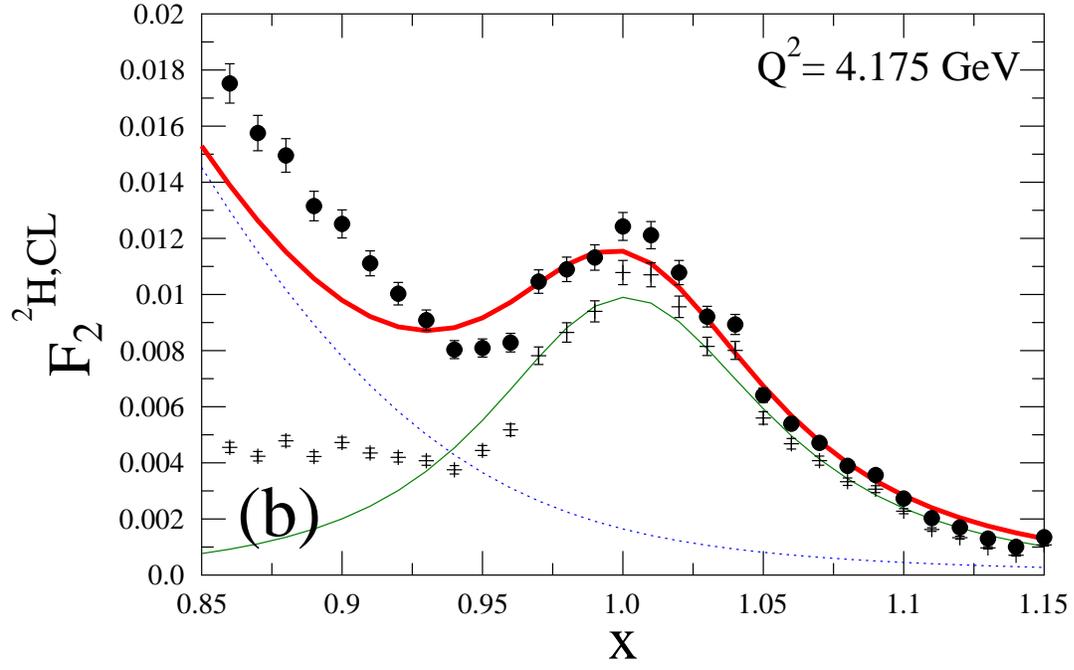}
\vspace{-12cm}\caption{ As Fig.10a for $Q^2=4.175$ GeV$^2$.}
\end{figure}

\renewcommand\thefigure{11}

\begin{figure}[p]
\includegraphics[scale=.8]{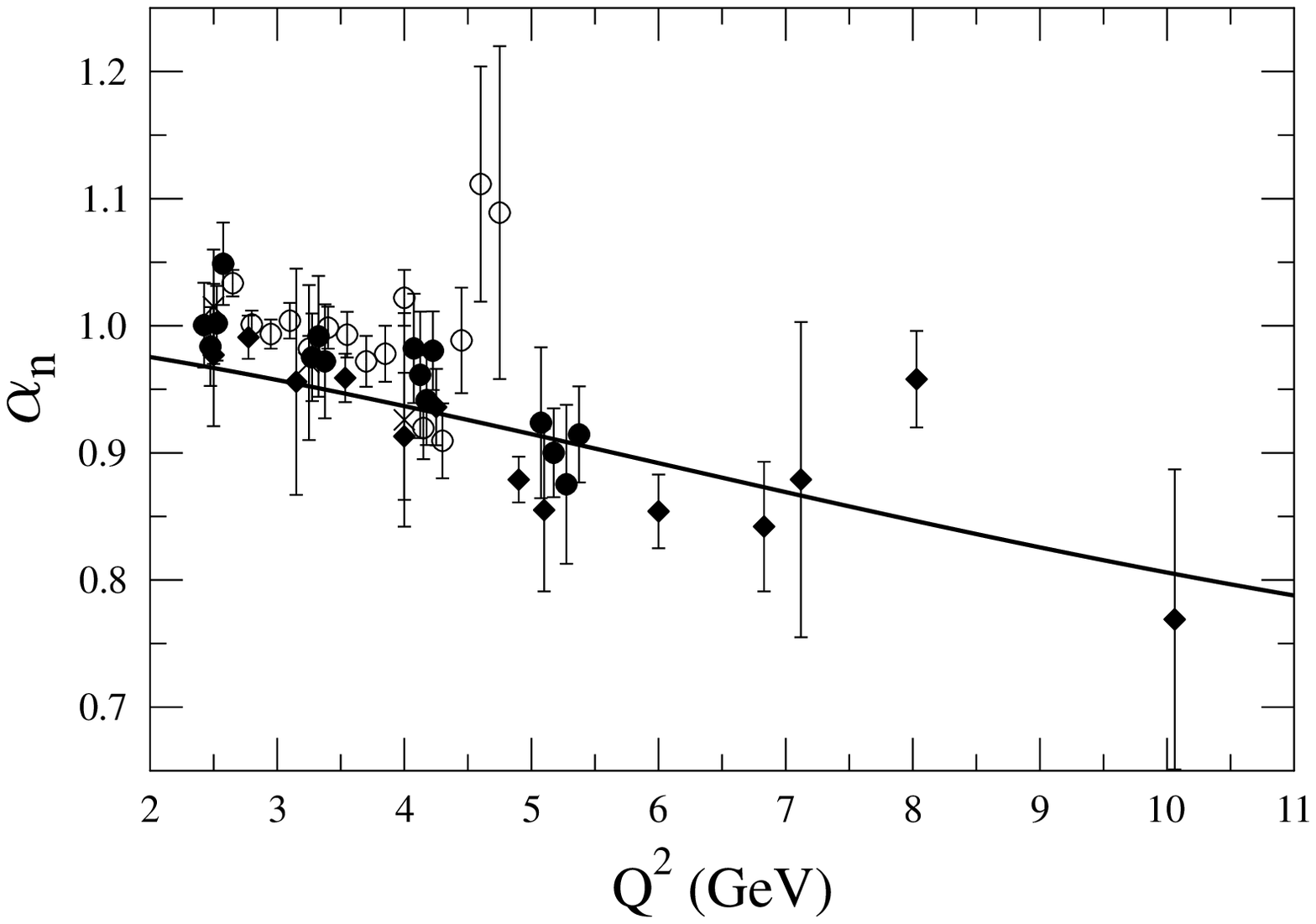}
\vspace{-12cm}\caption{Averaged reduced neutron magnetic $\alpha_n$
(Table II). On the curve for OD filled circles are results from CL
(Section III), crosses from Lung \cite{lung}, and diamonds for all
results from Table I in Ref. \cite{rtv} for $Q^2\ge 2.5$ GeV$^2$.
Open circles are from a $D(e,e'n)p/D(e,e'p)n$ experiment (Fig. 4 ,
Ref. \cite{brooks} . }
\end{figure}

\renewcommand\thefigure{12a}

\begin{figure}[p]
\includegraphics[scale=.8]{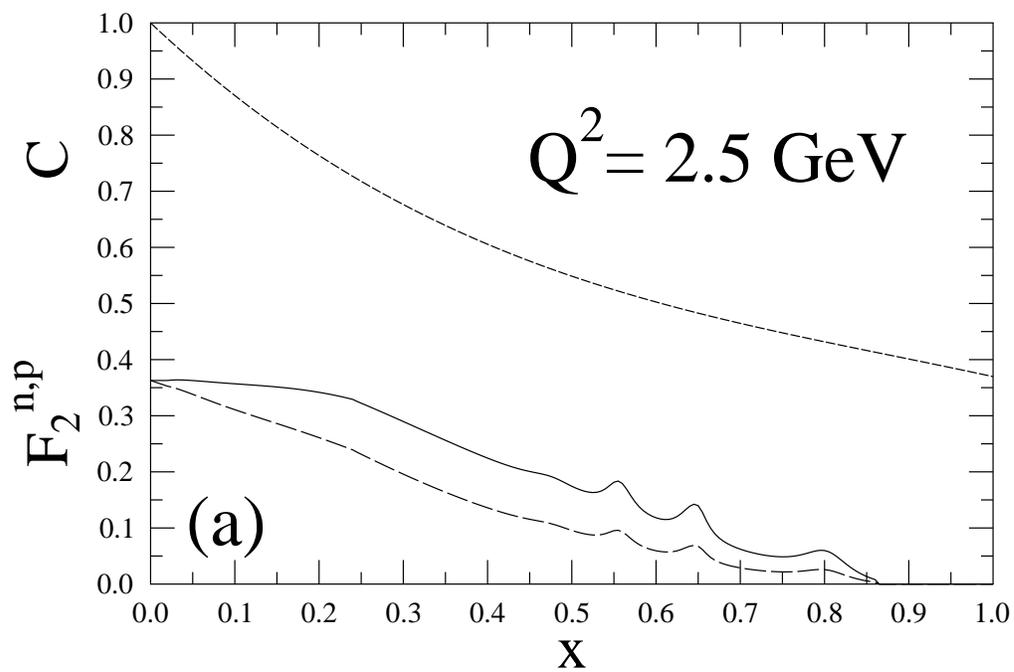}
\vspace{-12cm}\caption{Extracted $C(x,Q^2=2.5 \,{\rm GeV}^2)$ for
$x_M=0.75$ (short dashes) and $F_2^{p,n}(x,Q^2=2.5 \,{\rm GeV}^2)$
(drawn and dashed curves).}
\end{figure}

\renewcommand\thefigure{12b}

\begin{figure}[p]
\includegraphics[scale=.8]{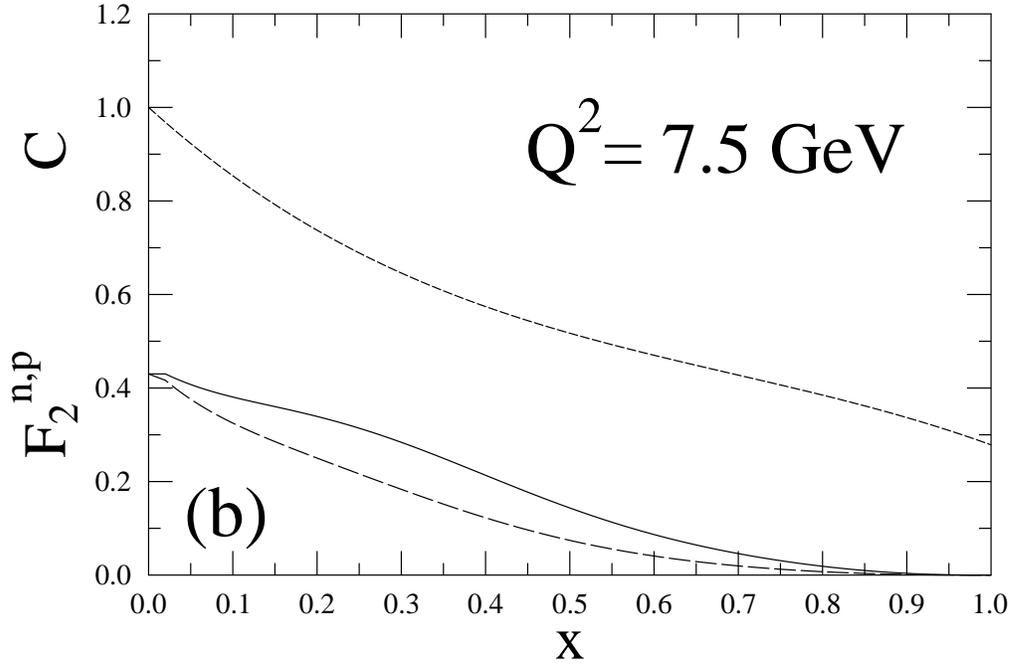}
\vspace{-12cm}\caption{As Fig. 12a for $Q^2=7.5 \,{\rm GeV}^2$.}
\end{figure}

\end{document}